\documentclass[11pt,reprint,superscriptaddress,aps,pre,nolongbibliography]{revtex4-2}
\usepackage[T1]{fontenc}
\usepackage{amsmath,amssymb}
\usepackage{booktabs}
\usepackage{array}
\usepackage{mathtools}
\usepackage{graphicx}
\usepackage{bm}
\usepackage{float}
\usepackage{comment}
\usepackage{cancel}
\usepackage{xcolor}
\usepackage{threeparttable}
\usepackage[caption=false]{subfig}
\usepackage[
  colorlinks=true,
  linkcolor=blue,
  citecolor=blue,
  urlcolor=blue,
  pdfborder={0 0 0}
]{hyperref}

\newcommand{\norm}[1]{\left\lVert #1 \right\rVert}
\newcommand{\abs}[1]{\left\lvert #1 \right\rvert}
\newcommand{\software}{RFBniCS}

\usepackage[all]{hypcap}

\usepackage[nameinlink,capitalise]{cleveref}
\crefname{equation}{Eq.}{Eqs.}
\Crefname{equation}{Equation}{Equations}
\crefname{paragraph}{paragraph}{paragraphs}
\Crefname{paragraph}{Paragraph}{Paragraphs}

\begin{document}

\raggedbottom

\title{\software{}: An open-source simulation framework for redox flow batteries}

\author{Amirhossein Aghabarari}
\affiliation{Department of Data Science, Norwegian University of Life Sciences, \AA{}s, Norway}

\author{Jørgen S. Dokken}
\affiliation{Simula Research Laboratory, Oslo, Norway}

\author{Martin T. Horsch}
\affiliation{Department of Data Science, Norwegian University of Life Sciences, \AA{}s, Norway}

\author{Eirik Valseth}
\affiliation{Simula Research Laboratory, Oslo, Norway}
\affiliation{Department of Mechanical Engineering and Technology Management, Norwegian University of Life Sciences, \AA{}s, Norway}

\author{Mathijs Janssen}
\affiliation{Department of Physics, Norwegian University of Life Sciences, \AA{}s, Norway}

\date{\today}

\begin{abstract}
    We present \software{}, an open-source finite-element framework implemented in FEniCSx for simulating redox-flow-battery (RFB) half-cells.
    \software{} solves an established macro-homogeneous porous-electrode model accounting for strongly coupled electrolyte flow, multicomponent species transport, ionic and electronic charge conservation, and interfacial Faradaic charge transfer. 
    Different from other open-access tools, \software{} can perform transient RFB half-cell simulations in one-, two-, and three-dimensional geometries while explicitly resolving the transport of both redox-active and supporting-electrolyte species. 
    It also describes electrolyte flow through porous electrodes and, in the three-dimensional formulation, through adjacent flow channels. 
    To verify the implementation, \software{}'s predictions in parameter-limiting regimes are compared against other published implementations, with \software{} generally showing superior accuracy and speed. 
    We further demonstrate \software{}'s capabilities by a simulation of the transient response of the negative half-cell of a vanadium redox flow battery, with a moderately concentrated supporting electrolyte. 
    In the current implementation, \software{} provides the computational basis for studying redox-flow-battery half-cells and can be extended to complex flow-field designs, alternative chemistries, full-cell coupling, detailed membrane transport, and additional multiphysics effects.
\end{abstract}

\maketitle

\section{Introduction}

\begin{figure}[t]
  \centering
  \includegraphics[width=1.0\linewidth]{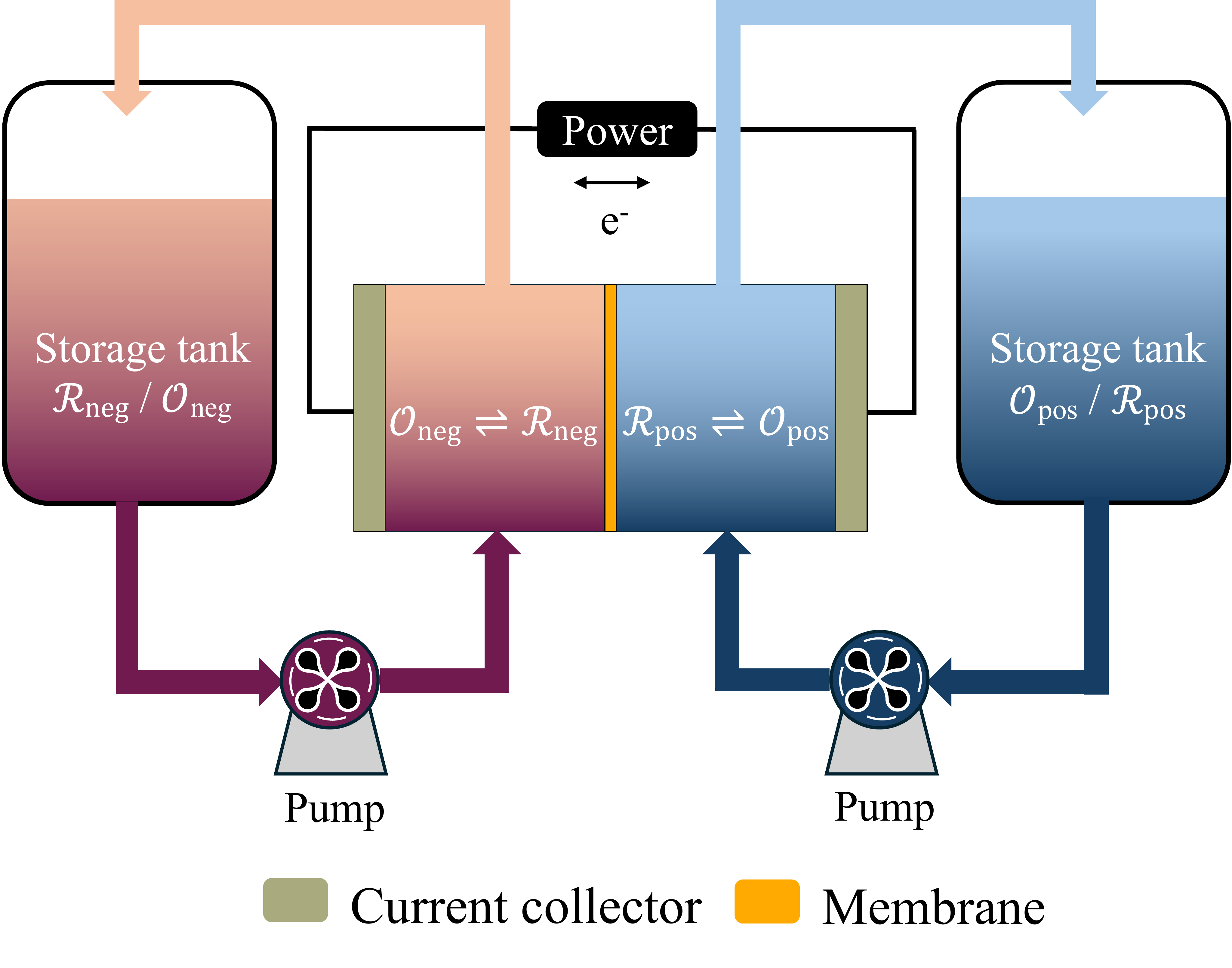}
  \caption{Schematic of a redox flow battery.}
  \label{fig:vrfb_schematic}
\end{figure}

To make full use of intermittent energy sources such as solar and wind, 
large-scale energy-storage technologies are needed~\cite{Aneke2016EnergyStorage}. 
Among the various alternatives, redox flow batteries (RFBs) stand out for their safety, cycle life, and scalability~\cite {Patel2024VRFBReliability,Beyes2025FlowBatteryTechnoEconomic,Puleston2022VRFBreview}.
RFBs store energy in redox-active species dissolved in liquid electrolytes, circulated from two storage tanks through two half-cells, separated by an ion-exchange membrane or separator~\cite{Li2017LowCostRFB,Sharmoukh2025RFBs}, see~\cref{fig:vrfb_schematic}.
Although RFBs generally have lower energy densities than many conventional battery systems, this limitation is less restrictive for stationary applications, where their energy capacity can be scaled by increasing the electrolyte tank volume and, within solubility limits, the concentration of redox-active species~\cite{Puleston2022VRFBreview}.
Several RFB chemistries have been developed, including all-vanadium systems, commonly referred to as vanadium redox flow batteries (VRFBs), iron--chromium systems, zinc--bromine systems, and organic redox-active materials~\cite{Leung2017OrganicRFBReview,Liu2023ProgressRFB}.

An RFB half-cell consists of a porous electrode, usually made from fibrous carbon felt with a high volumetric surface area, positioned between a current collector and an ion-exchange membrane or separator~\cite{Ghimire2021}.
Electrolytes are pumped through these porous electrodes to supply redox-active species.
During charging, the redox-active species react at the solid--electrolyte interface according to the half-cell reactions~\cite{Olabi2023RFBReview,Bard2001ElectrochemicalMethods}
\begin{subequations}
\begin{align}
\text{positive half-cell:} \qquad
& \mathcal{R}_{\mathrm{pos}}
\rightarrow
\mathcal{O}_{\mathrm{pos}} + n_e\,\mathrm{e}^{-},
\\
\text{negative half-cell:} \qquad
& \mathcal{O}_{\mathrm{neg}} + n_e\,\mathrm{e}^{-}
\rightarrow
\mathcal{R}_{\mathrm{neg}}\label{eq:negative_half_cell}.
\end{align}
\end{subequations}
Here, \(\mathcal{R}\) denotes the reduced and \(\mathcal{O}\) the oxidized form of the redox species, and \(n_e\) is the number of electrons, $\mathrm{e}^{-}$, transferred per redox reaction (--).
During discharge, these reactions reverse.
By using porous electrodes, Faradaic reactions occur over a large interfacial area, reducing the local Faradaic current density and thereby lowering activation overpotentials~\cite{HAVERKORT2024}.
Moreover, in porous electrodes, redox-active species only need to move a short distance from the bulk electrolyte to the solid–electrolyte interface~\cite{LI20124634,KE2015109}. 
This helps maintain a steady supply of reactants and minimizes mass-transport limitations.
In addition to the redox-active species, RFB electrolytes usually contain several supporting-electrolyte species, whose high concentrations yield a high ionic conductivity, damping the electric field and minimizing Ohmic losses.

The flow-channel geometry in RFBs controls how the electrolyte is distributed through the porous electrode, influencing pressure drop, reactant availability, local Faradaic reaction rates, and hence RFB performance~\cite{Hu2025,Peng2024FlowFieldVisualization}.
Several flow-field configurations are used in RFBs, including interdigitated, parallel, serpentine, and flow-through flow fields (FTFF)~\cite{Chakrabarti2020RFBModellingReview,Sun2023StackScaleFlowFields,Yadav2023PressureDrop,Yu2025,Liu2026}.
RFB performance, including its efficiency, power density, reactant utilization, and pressure drop, is further influenced by its porous-electrode microstructure~\cite{Hu2025,Barzegari2025TopologyPorous}, membrane selectivity~\cite{Xue2026}, parasitic side reactions~\cite{Ma2024}, operating temperature~\cite{Bogdanov2026}, and other structural and operational parameters~\cite{Talebian2025,Vishnu2026}.
Experimental optimization is thus costly and time-consuming, further complicated by the fact that transport and reaction processes inside RFBs are difficult to observe experimentally.
Mathematical models can help interpret internal transport and reaction phenomena, understand performance limitations, and optimize RFB designs.

RFB models cover a range of length scales and resolutions~\cite{Chakrabarti2020RFBModellingReview}.
At one end of the spectrum, lumped models represent RFBs using spatially averaged variables, allowing for rapid simulations and system-level performance evaluation.
On the other end of the spectrum, pore-scale models explicitly resolve the local concentration, potential, velocity fields, and interfacial Faradaic reactions within realistic porous-electrode microstructures---though, due to their high computational cost, only on small domains~\cite{Liu2025RecentAdvancesARFBModeling,Qiu2012PoreScale,Hunt2020EffectiveThermal}.
In between are macro-homogeneous porous-electrode models. 
Such models describe transport and reaction processes in porous electrodes using a volume-averaged continuum formulation, in which pore-scale properties are represented through effective parameters such as porosity, tortuosity, permeability, and specific interfacial area~\cite{NEWMAN2021,XU2019115864}.
The solid and electrolyte phases are treated as interpenetrating phases that coexist at the macroscopic scale, coupled through interfacial Faradaic source terms~\cite{HAVERKORT2024,NEWMAN197525}.
Macro-homogeneous porous-electrode models retain the key physics required to describe macroscopic cell behavior, including the conservation of mass, momentum, and charge, while parametric studies and design analyses remain computationally feasible.
They provide spatially resolved distributions of internal quantities, such as species concentrations, overpotential, current density, and flow fields, which are essential for practical design studies. 
Membrane transport, side reactions, thermal effects, and water distribution can be incorporated as well.

Many RFB simulations based on macro-homogeneous porous-electrode models have been performed with commercial software~\cite{SHAH20088087,ALI2025237659,LI20255192,AbdelMessih2025VRFBthermal,WANG2024Numerical,MartinezLopez2024_2D_VRFB,MartinezLopez2025_Obstacle_ORFB,Xiao2025_DES_RFB,DAdamo2025_NAORFB_COMSOL,BROMBERGER201464}.
As an alternative, open-source tools have distinct benefits: besides being cost-effective, access to the source code enables users to inspect, customize, and extend open-source software to meet specific modeling needs. 
Several open-source simulation frameworks have been developed for lithium-ion batteries, including PyBaMM~\cite{Sulzer2021}, BattMo~\cite{Raynaud2025BattMo}, LIONSIMBA~\cite{Torchio2016LIONSIMBA}, PETLION~\cite{berliner2021petlion}, and COBRAPRO~\cite{cobrapro_jes_2024,cobrapro_joss_2025}.
Several open-source tools for RFBs were developed recently as well, including RFBzero~\cite{Fell2024RFBzero}, EZBattery~\cite{BAO2025118134}, and RfbFoam~\cite{Barzegari2026RfbFoam}.
These RFB frameworks vary in their modeling assumptions, governing equations, boundary conditions, dimensionality, geometric representations, and simulation outputs.
RFBzero is a zero-dimensional Python framework for simulating full- and symmetric-cell cycling, with optional descriptions of redox-active-species degradation and crossover.
EZBattery is an analytical framework with selectable zero- and two-dimensional schemes for predicting full-cell charge–discharge voltage curves; its two-dimensional scheme represents porous electrodes as idealized microchannels.
RfbFoam is a framework for macro-homogeneous RFB modeling, built on OpenFOAM~\cite{greenshields2025}.
The framework is formulated for half-cell simulations and solves the steady-state coupled momentum, species, and charge transport equations in three-dimensional geometries in the excess supporting electrolyte limit.
To the best of the authors’ knowledge, there is no open-access RFB tool with peer-reviewed or sufficiently detailed documentation that combines spatially resolved transient simulations with explicit multicomponent electrolyte transport beyond the excess-supporting-electrolyte approximation.

To address this need, we developed \software{}, an open-source framework for RFB modeling, beyond steady-state and excess supporting electrolyte assumptions.
The model is implemented in FEniCSx~\cite{DOLFINx2023}, an open-source computing environment for solving partial differential equations using the finite element method. 
In a fully coupled finite-element formulation, \software{} solves a well-established macro-homogeneous porous-electrode model accounting for interfacial Faradaic reactions, ionic and electronic charge conservation coupled to species transport through diffusion, electromigration, and advection. 
In the present implementation, \software{} focuses on an RFB half-cell and treats the membrane as an interface. 
This setup provides a basis for developing the numerical framework, evaluating its performance, and testing it under realistic conditions. 
A full-cell model will be addressed in future work, as it requires the two half-cells, the thin membrane, and the transport and electrochemical coupling between these regions to be treated consistently.
While we highlight \software{}'s performance here for a VRFB half-cell, the open-source implementation allows users to inspect, modify, and extend the model, including the geometry, transport-property models, boundary conditions, kinetic expressions, redox chemistry, and other model components.
In particular, modifying the governing equations allows different physical phenomena to be included or excluded depending on the desired level of model complexity.

This paper is organized as follows.
\Cref{sec:model} presents the model setup, governing equations, and boundary conditions. 
\Cref{sec:methods} describes the numerical implementation, weak formulations, sequential flow--electrochemical solution strategy, and time discretization. 
\Cref{sec:results} assesses the model through verification and comparison cases for selected limiting regimes and presents representative simulation results. 
Finally, \cref{sec:conclusions} summarizes the main findings and outlines possible extensions of the framework.

\section{Model}\label{sec:model}

\subsection{Modeling domain}

\software{} includes one-, two-, and three-dimensional variants.
All variants use the macro-homogeneous porous-electrode framework described in \cref{sec:governing_equations}, but differ in their computational domains, flow models, and boundary conditions.
In this way, users can select an appropriate level of model fidelity based on their study objectives, required accuracy, and available computational resources.
The computational domain is denoted by \(\Omega\subset\mathbb{R}^{n_{\mathrm{dim}}}\), where \(n_{\mathrm{dim}}=1\), \(2\), or \(3\). 
Its external boundary is denoted by \(\partial\Omega\), and the outward unit normal vector on \(\partial\Omega\) is denoted by \(\bm n\).

\begin{figure}
  \centering
    \subfloat[\label{fig:model_1d}]{
        \includegraphics[width=\linewidth]
        {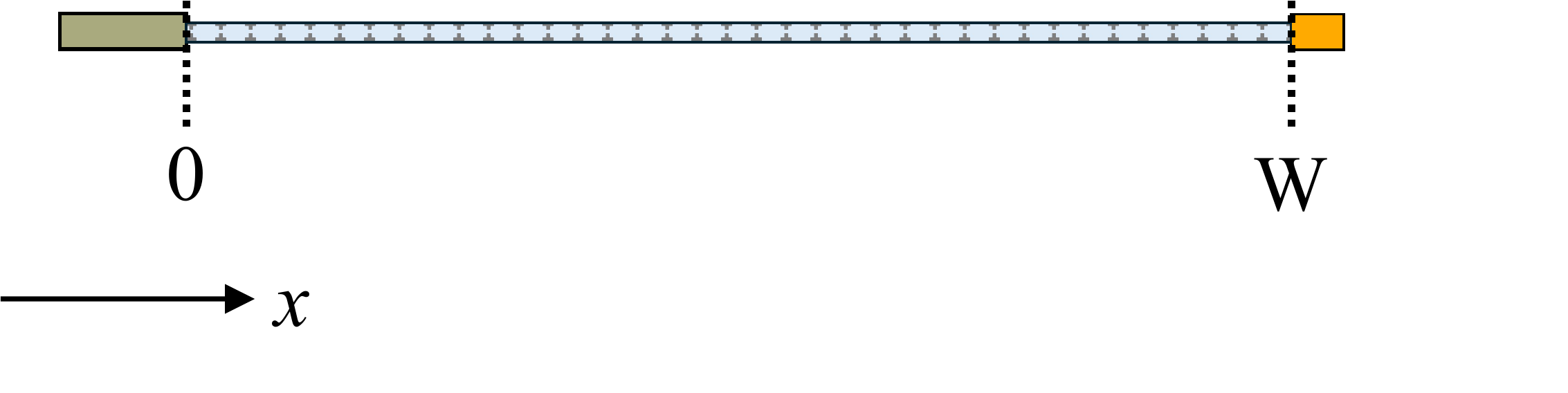}
    }
    \par\medskip
    
    \subfloat[\label{fig:model_2d}]{
        \includegraphics[width=\linewidth]
        {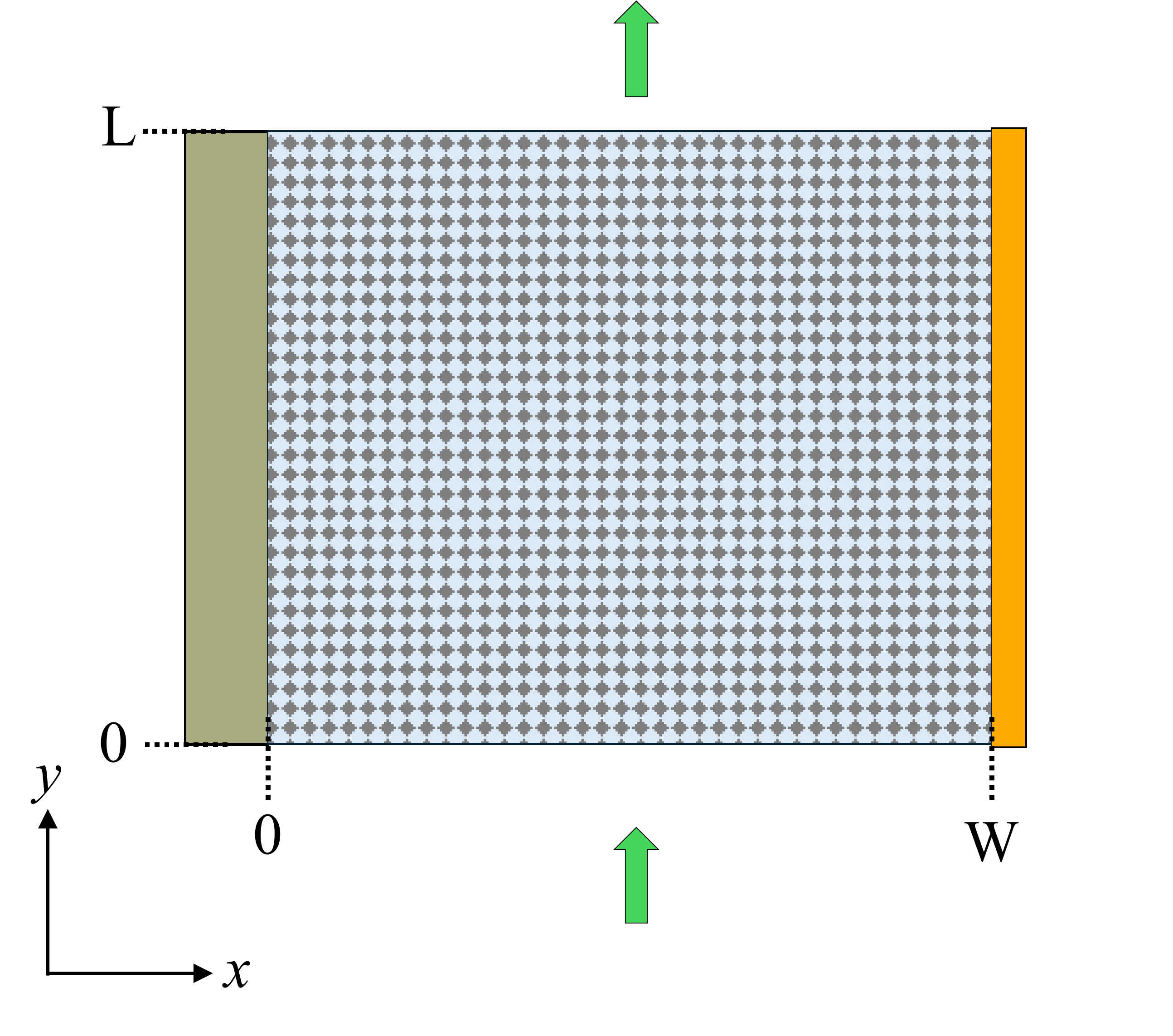}
    }
    \par\medskip
    
    \subfloat[\label{fig:model_3d}]{
        \includegraphics[width=\linewidth]
        {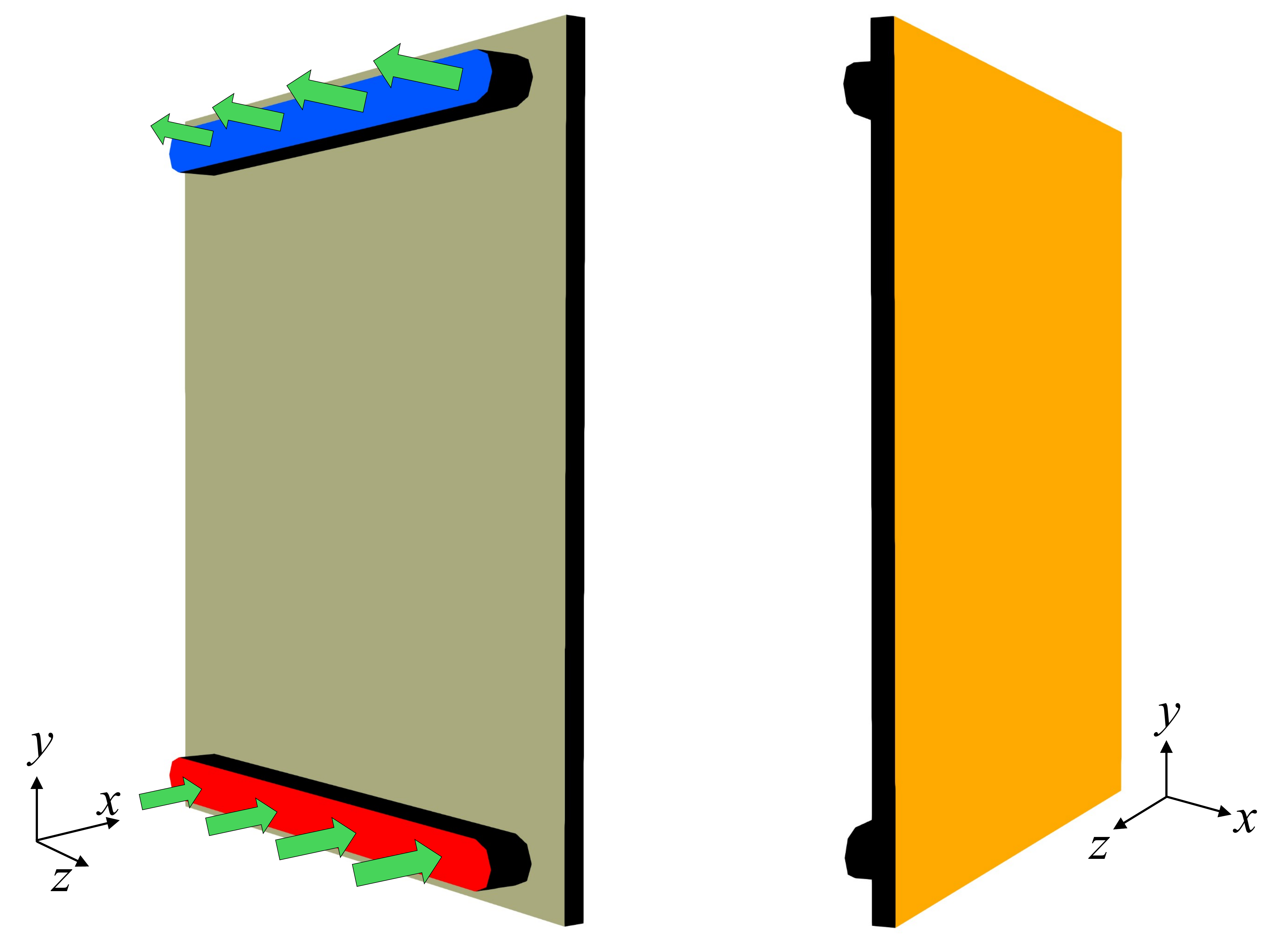}
    }

    \par\medskip

    \includegraphics[width=\linewidth]
    {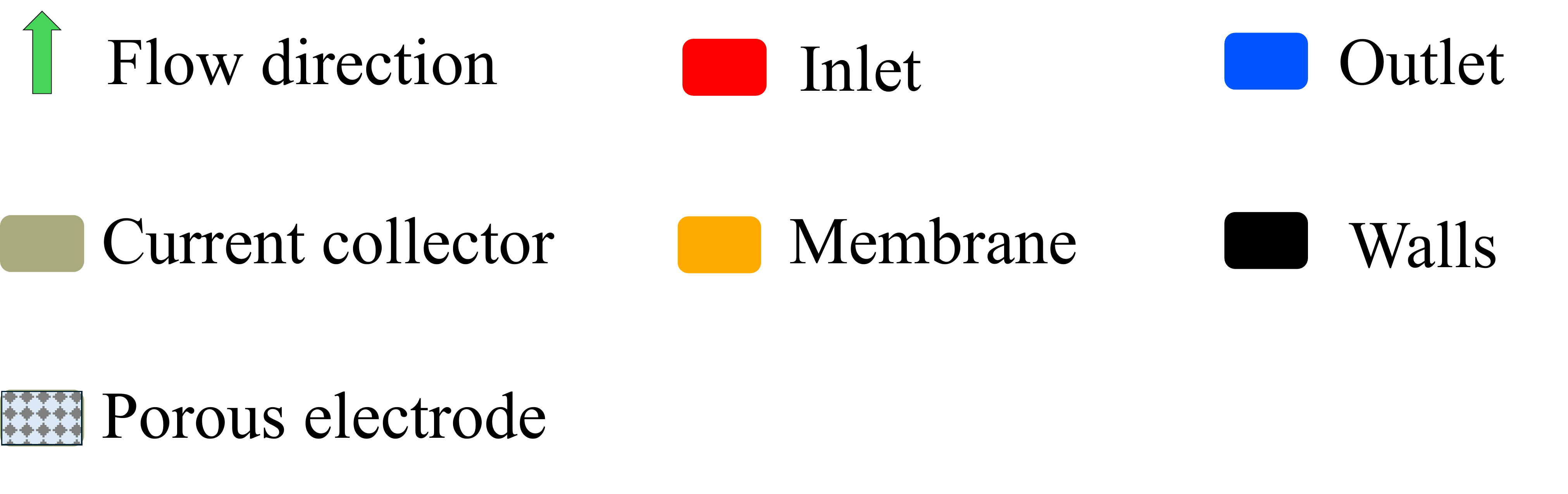}
    
  \caption{Schematics of the (a) one-dimensional, (b) two-dimensional, and (c) three-dimensional half-cell domains. 
  In panel (c), the same domain is shown from two different viewing angles to clarify its geometry and boundary arrangement. The legend applies to all three domains.}

  \label{fig:model_2d_and_3d}
\end{figure}

The one-dimensional (1D) model represents a porous electrode of width $W$ with a coordinate $x$ running from the current collector at $x=0$ to the membrane at $x=W$, see \cref{fig:model_2d_and_3d}\mbox{(a)}. 
Accordingly, the 1D computational domain is defined as $\Omega=(0,W)\subset\mathbb{R}$.
The 1D model does not resolve electrolyte flow, and the advective contribution to species transport is therefore omitted.
Its low computational cost makes it well-suited for rapid parametric studies.

In the two-dimensional (2D) variant, the electrolyte enters the porous electrode of width $W$ and length $L$ at $y=0$ and leaves it at $y=L$, see \cref{fig:model_2d_and_3d}\mbox{(b)}.
The porous electrode faces a current collector at \(x=0\) and a membrane at \(x=W\).
For this geometry the computational domain is $\Omega=(0,W)\times(0,L)\subset\mathbb{R}^{2}$.
Unlike the 1D model, the 2D model can capture variations along the flow direction and describe coupled advection and electrochemical processes.
Such 2D porous-electrode formulations, which resolve the through-plane and flow directions while assuming invariance in the out-of-plane direction, are commonly used in RFB modeling as a compromise between physical resolution and computational cost~\cite{SHAH20088087,BROMBERGER201464,MartinezLopez2024_2D_VRFB}.

In the three-dimensional (3D) version of the model, we consider an FTFF configuration as shown in~\cref{fig:model_2d_and_3d}\mbox{(c)}.
This model contains flow-channel domains beside the porous electrode. 
The electrolyte enters the computational domain through the inlet channel, passes through the porous electrode, and exits through the outlet channel, as indicated by the flow arrows. 
The porous electrode is not visible in \cref{fig:model_2d_and_3d}\mbox{(c)} because it is enclosed by the surrounding walls, current-collector, and membrane interfaces.
We decompose the domain into cell-wise disjoint subregions, $\Omega=\Omega_E\cup\Omega_F$, $\Omega_E\cap\Omega_F=\emptyset$, where \(\Omega_E\) denotes the porous-electrode region, and \(\Omega_F\) denotes the non-porous flow-channel region. 
For the 1D and 2D porous-electrode models, the entire computational domain is the porous-electrode region, so that \(\Omega_E=\Omega\).
The 3D model enables a detailed analysis of the effects of channel geometry, inlet/outlet arrangement, and flow distribution on half-cell performance, not possible in 1D or 2D formulations.

Different portions of the boundary are denoted by \(\Gamma\).
In particular, \(\Gamma_{\mathrm{cc}}\) and \(\Gamma_{\mathrm{mem}}\) denote the current-collector and membrane boundaries, respectively. 
For models in which the electrolyte flow is resolved, the inlet and outlet are denoted by \(\Gamma_{\mathrm{in}}\) and \(\Gamma_{\mathrm{out}}\), respectively.
The 3D model also includes the non-conductive walls, grouped under \(\Gamma_{\mathrm{wall}}\). 
The external boundary can therefore be written as $\partial\Omega = \Gamma_{\mathrm{cc}} \cup \Gamma_{\mathrm{mem}} \cup \Gamma_{\mathrm{in}} \cup \Gamma_{\mathrm{out}} \cup \Gamma_{\mathrm{wall}}$, with only the boundary segments relevant to each model dimension being included.

\subsection{Governing equations}\label{sec:governing_equations} 

\subsubsection{Species conservation and fluxes}

Let \(\mathcal{J}\) denote the set of ionic species in the electrolyte of an RFB half-cell.
\(\mathcal{J}\) consists of a reduced species $\mathcal{R}$, an oxidized species $\mathcal{O}$, and several supporting-electrolyte species.
In volume-averaged form, conservation of species \(j\in\mathcal{J}\) in the porous electrode gives~\cite{NEWMAN2021}
\begin{align}
\frac{\partial \!\left(\varepsilon c_j\right)}{\partial t}
\;+\; \nabla \!\cdot \bm{N}_j
&= S_j,
\label{eq:mass_balance}
\end{align}
where $\varepsilon$ is the porous-electrode porosity (--), $c_j$ denotes the local concentration of species $j$ in the electrolyte ($\mathrm{mol\,m^{-3}}$), $\bm{N}_j$ is the corresponding molar flux ($\mathrm{mol\,m^{-2}\,s^{-1}}$), and $S_j$ is a volumetric source term arising from interfacial Faradaic reactions ($\mathrm{mol\,m^{-3}\,s^{-1}}$).

Assuming local electroneutrality, the ionic concentrations satisfy
\begin{equation}
\sum_{j\in\mathcal{J}} z_j c_j = 0,
\label{eq:electroneutrality}
\end{equation}
where \(z_j\) is the charge number of species \(j\) (--). 
Due to \cref{eq:electroneutrality}, only \(\lvert\mathcal{J}\rvert-1\) ionic concentrations are independent.
One of the supporting-electrolyte ions, denoted by \(d\in\mathcal{J}\), is selected as the dependent species.
One can then solve \cref{eq:mass_balance} only for \(j\in\mathcal{J}\setminus\{d\}\), and find $c_d$  by \cref{eq:electroneutrality}.

The molar flux of species \(j\) is described using a Nernst--Planck equation, modified for transport in porous media~\cite{NEWMAN2021}, accounting for diffusion, electromigration, and advection,
\begin{equation}
\bm{N}_j= -D_j^{\mathrm{eff}}\nabla c_j 
            - \frac{z_j c_j D_j^{\mathrm{eff}}}{R T} F \nabla \phi_\ell 
            + \bm{v}\, c_j,
\label{eq:Nernst-Planck}
\end{equation}
where $F$ is Faraday's constant ($\mathrm{C\,mol^{-1}}$), $R$ is the universal gas constant ($\mathrm{J\,mol^{-1}\,K^{-1}}$), $T$ is the absolute temperature ($\mathrm{K}$), and $\phi_\ell$ is the electrolyte-phase electric potential ($\mathrm{V}$), with the subscript ($\ell$) denoting the liquid phase.
The superficial (Darcy) electrolyte velocity $\bm{v}$ ($\mathrm{m\,s^{-1}}$) is related to the interstitial (pore) average velocity $\bm{u}$ ($\mathrm{m\,s^{-1}}$) by $\bm{u}=\bm{v}/\varepsilon$.
The effective diffusivity $D_j^{\mathrm{eff}}$  of species $j$ ($\mathrm{m^2\,s^{-1}}$) depends on the porosity and tortuosity of a  porous electrode; we employ a  Bruggeman-type approximation \cite{TJADEN201644}, 
\begin{equation}
    D_j^{\mathrm{eff}}=\varepsilon^{3/2}D_j,
    \label{eq:Bruggeman}
\end{equation}
where $D_j$ is the molecular diffusivity of species $j$ in the bulk electrolyte.

For a single-electron charge transfer, the source terms follow from Faraday’s law as
\begin{equation}
S_j =
\begin{cases}
-\dfrac{a i_F}{F}, 
& j = \mathcal{R},\\[6pt]
+\dfrac{a i_F}{F}, 
& j = \mathcal{O},\\[6pt]
0, 
& j\in\mathcal{J}\setminus\{\mathcal{R},\mathcal{O}\},
\end{cases}
\label{eq:species_source_terms}
\end{equation}
where \(a\) is the specific interfacial area \((\mathrm{m^2\,m^{-3}})\), and \(i_F\) is the interfacial Faradaic current density \((\mathrm{A\,m^{-2}})\).
We use the anodic-positive sign convention: \(i_F>0\) denotes oxidation and \(i_F<0\) reduction.

\subsubsection{Electrolyte-phase ionic current density}
The ionic current density \((\mathrm{A\,m^{-2}})\) is obtained by 
\begin{equation}
\bm{i}_\ell = \sum_{j\in\mathcal{J}} \bm{i}_j,
\label{eq:current_density}
\end{equation}
where $\bm{i}_j = z_j F \bm{N}_j$ is the current density carried by species \(j\)~\cite{NEWMAN2021}.
Using this relation together with \cref{eq:Nernst-Planck} in \cref{eq:current_density} yields
\begin{equation}
\bm{i}_\ell
= -\kappa^{\mathrm{eff}} \nabla \phi_\ell
  - F\sum_{j\in\mathcal{J}} z_j D_j^{\mathrm{eff}} \nabla c_j,
\label{eq:electrolyte-current-density-compact}
\end{equation}
where the advective contribution vanishes by local electroneutrality and 
\begin{equation}
\kappa^{\mathrm{eff}}
= \frac{F^2}{RT}\sum_{j\in\mathcal{J}} z_j^2 D_j^{\mathrm{eff}} c_j, 
\label{eq:effective-conductivity}
\end{equation}
is the effective ionic conductivity \((\mathrm{S\,m^{-1}})\).

\subsubsection{Solid-phase electronic current density}
We use Ohm's law for the electronic current density in the solid phase of the porous electrode (\(\mathrm{A\,m^{-2}}\)), 
\begin{equation}
\bm{i}_{s} = -\sigma^{\mathrm{eff}} \nabla \phi_s,
\label{eq:solid_ohms_law}
\end{equation}
where \(\phi_s\) is the electric potential in the solid phase \((\mathrm{V})\), and \(\sigma^{\mathrm{eff}}\) is the effective electronic conductivity of the solid phase within the porous electrode \((\mathrm{S\,m^{-1}})\). We estimate the latter through the Bruggeman relation \(\sigma^{\mathrm{eff}} = (1-\varepsilon)^{3/2}\sigma\), where \(\sigma\) is the intrinsic conductivity of the solid phase \((\mathrm{S\,m^{-1}})\).

\subsubsection{Interfacial Faradaic charge-transfer kinetics}

Interfacial Faradaic reactions transfer charge between the electrolyte and solid phases within the porous electrode~\cite{HAVERKORT2024}.
By multiplying \cref{eq:mass_balance} by \(z_jF\), summing over all ionic species, and using local electroneutrality, the definition of \(\bm{i}_\ell\), and the source terms in \cref{eq:species_source_terms}, we obtain
\begin{equation}
\nabla\cdot\bm{i}_{\ell}=a\,i_F.
\label{eq:electrolyte_charge_balance}
\end{equation}
Conservation of total current,
\(\nabla\cdot(\bm{i}_{\ell}+\bm{i}_{s})=0\), then gives
\begin{equation}
\nabla\cdot\bm{i}_{s}=-a\,i_F.
\label{eq:solid_charge_balance}
\end{equation}

The interfacial Faradaic current density is modeled by the Butler--Volmer relation \cite{MunozPerales2022OperationVariablesVRFB, Murthy2018ConcentrationOverpotentialVRFB},
\begin{equation}
i_F = i_0
\left[
\frac{c_{\mathcal{R}}^{\mathrm{surf}}}{c_{\mathcal{R}}}
\exp\!\left(\frac{\alpha_a F}{RT}\,\eta\right)
-
\frac{c_{\mathcal{O}}^{\mathrm{surf}}}{c_{\mathcal{O}}}
\exp\!\left(-\frac{\alpha_c F}{RT}\,\eta\right)
\right],
\label{eq:butler_volmer}
\end{equation}
where the exchange current density \(i_0\) is defined as
\begin{equation}
 i_0 = F k^{0}
 \left(c_{\mathcal{R}}\right)^{\alpha_c}
 \left(c_{\mathcal{O}}\right)^{\alpha_a},
\label{eq:exchange_current_density}
\end{equation}
with \(k^{0}\) being the standard rate constant \((\mathrm{m\,s^{-1}})\) and \(\alpha_a\) and \(\alpha_c\) the anodic and cathodic transfer coefficients (--).
In \cref{eq:butler_volmer}, the activation overpotential \(\eta\) \((\mathrm{V})\) is defined as
\begin{equation}
  \eta = \phi_s - \phi_\ell - E_{\mathrm{eq}},
  \label{eq:overpotential}
\end{equation}
where the equilibrium potential \(E_{\mathrm{eq}}\) \((\mathrm{V})\) follows Nernst's equation,
\begin{equation}
E_{\mathrm{eq}} = E_0' - \frac{RT}{F}
\ln\!\left( \frac{c_{\mathcal{R}}}{c_{\mathcal{O}}} \right),
\label{eq:Eeq}
\end{equation}
and \(E_0'\) is the formal potential of the redox couple \((\mathrm{V})\).

The Butler--Volmer equation \eqref{eq:butler_volmer} contains the interfacial concentrations at the solid--electrolyte interface $c_\mathcal{R}^{\mathrm{surf}}$ and $c_\mathcal{O}^{\mathrm{surf}}$ of the reduced and oxidized species.
These may differ from the corresponding bulk concentrations because Faradaic reactions consume or produce species at the solid--electrolyte interface, requiring transport from the bulk electrolyte to the interface for consumed species and in the opposite direction for produced species.
Following the approach in Ref.~\cite{MunozPerales2022OperationVariablesVRFB}, we use a linear mass-transfer relation, 
\begin{equation}
    N_j^{\mathrm{surf}} = k_{j}\left(c_j-c_j^{\mathrm{surf}}\right),
    \qquad j\in\{\mathcal{R},\mathcal{O}\},
\end{equation}
for the molar flux \(N_j^{\mathrm{surf}}\) of species \(j\), from the bulk electrolyte to the solid--electrolyte interface---not to be confused with the bulk molar-flux vector \(\bm{N}_j\).
Following Ref.~\cite{KRISHNAMURTHY201110019}, we use a semi-empirical correlation for the mass-transfer coefficient \(k_{j}\) \((\mathrm{m\,s^{-1}})\),
\begin{equation}
    k_{j}
    =
    7\frac{D_j}{d_f}\mathrm{Re}^{0.4},
    \qquad j \in \{\mathcal{R},\mathcal{O}\},
    \label{eq:mass_transfer_km}
\end{equation}
where \(d_f\) is the fiber diameter \((\mathrm{m})\), and \(\mathrm{Re}=\rho\norm{\bm{v}}d_f/\mu\) is the Reynolds number (--), based on the superficial electrolyte velocity \(\bm{v}\).
At the solid--electrolyte interface, the interfacial mass-transfer flux must balance the consumption or production of each species by the Faradaic reaction, giving $N_j^{\mathrm{surf}}=\pm i_F/(F)$; hence, 
\begin{equation}
    k_{j}\left(c_j - c_j^{\mathrm{surf}}\right) = \pm \frac{i_F}{F},
    \label{eq:interfacial_mass_transfer_balance}
\end{equation}
where the sign depends on whether species $j$ is consumed or produced at the solid--electrolyte interface.
\Cref{eq:interfacial_mass_transfer_balance} shows that the difference between bulk and surface concentrations is governed by a competition between the Faradaic reaction and mass transport.

As we show in Appendix~\ref{app:surface-concentrations}, combining the
interfacial mass-transfer balance relation~\eqref{eq:interfacial_mass_transfer_balance} with the Butler--Volmer equation~\eqref{eq:butler_volmer} yields interfacial concentrations consistent with Ref.~\cite{MunozPerales2022OperationVariablesVRFB}, 
\begin{subequations}\label{eq:surface_concentration}
\begin{align}
c_{\mathcal{R}}^{\mathrm{surf}} &=
\frac{c_\mathcal{R}\left(1+B/k_{\mathcal{O}}\right)+c_\mathcal{O}B/k_{\mathcal{R}}}
{1+A/k_{\mathcal{R}}+B/k_{\mathcal{O}}},\\
c_{\mathcal{O}}^{\mathrm{surf}} &=
\frac{c_\mathcal{O}\left(1+A/k_{\mathcal{R}}\right)+c_\mathcal{R}A/k_{\mathcal{O}}}
{1+A/k_{\mathcal{R}}+B/k_{\mathcal{O}}},\\
\intertext{where} 
A &= k^{0}\left(c_\mathcal{R}\right)^{\alpha_c-1}\left(c_\mathcal{O}\right)^{\alpha_a}
\exp\!\left(\frac{\alpha_a F}{RT}\eta\right),
\\[6pt]
B &= k^{0}\left(c_\mathcal{R}\right)^{\alpha_c}\left(c_\mathcal{O}\right)^{\alpha_a-1}
\exp\!\left(-\frac{\alpha_c F}{RT}\eta\right).
\end{align}
\end{subequations}
The mass-transfer correlation \eqref{eq:mass_transfer_km} and the resulting surface-concentration expressions \eqref{eq:surface_concentration} are retained only for the flow-resolved formulations. Thus, for the 1D formulation, the locally resolved electrolyte concentrations are used directly in the Butler--Volmer equation.

\subsubsection{Electrolyte flow}
We use different flow models depending on the dimensionality of the domain.
As noted previously, the 1D model omits electrolyte flow and thus the advective term in \cref{eq:Nernst-Planck}.
For the 3D model, we describe electrolyte flow throughout the computational domain using a unified volume-averaged Navier--Stokes formulation. 
In the porous-electrode region, \(\Omega_E\), additional Darcy and Forchheimer resistance terms are included~\cite{Whitaker1996Forchheimer,Barzegari2026RfbFoam}.
In conservative form, the governing equations are
\begin{subequations}\label{eq:NavierStokes}
\begin{align}
&\frac{\partial}{\partial t}
\left( \frac{\rho}{\varepsilon}\bm{v} \right)
+
\nabla\!\cdot
\left(
\frac{\rho}{\varepsilon^2}\bm{v}\otimes\bm{v}
\right)
\notag\\
&\quad=
-\nabla p
+
\nabla\!\cdot
\left(
\frac{\mu}{\varepsilon}\nabla\bm{v}
\right)
-
\frac{\mu}{K}\bm{v}
-
\rho\beta |\bm{v}|^\alpha \bm{v},
\label{eq:momentum_nsdf_general}
\\
&\frac{\partial(\varepsilon\rho)}{\partial t}
+
\nabla\!\cdot(\rho\bm{v})
=0 .
\label{eq:continuity_general}
\end{align}
\end{subequations}
Here, \(\rho\) is the electrolyte density \((\mathrm{kg\,m^{-3}})\), \(p\) is the liquid pressure \((\mathrm{Pa})\), \(\mu\) is the dynamic viscosity \((\mathrm{Pa\,s})\), \(K\) is the permeability of the porous electrode \((\mathrm{m^2})\), \(\alpha\) is the Forchheimer exponent (--), and \(\beta\) is the Forchheimer coefficient \((\mathrm{m^{-\alpha}\,s^{\alpha-1}})\).
\Cref{eq:momentum_nsdf_general} is the fluid momentum balance, in which the linear term \(-\mu\bm{v}/K\) represents Darcy resistance and the nonlinear term \(-\rho\beta |\bm{v}|^\alpha\bm{v}\) represents Forchheimer inertial resistance.
\Cref{eq:continuity_general} expresses fluid mass conservation in the porous medium. 

We estimate the permeability of the fibrous porous electrode using the Kozeny--Carman-type relation, \(K=d_f^{\,2}\varepsilon^{3}/[C_{\mathrm{KC}}(1-\varepsilon)^2]\),
where \(C_{\mathrm{KC}}\) is the Kozeny--Carman constant (--)~\cite{SHAH20088087}.
The Forchheimer coefficient is estimated using the Ergun-type relation \(\beta=1.75(1-\varepsilon)/(L_p\varepsilon^3)\), where \(L_p\) is the effective characteristic microstructural length \((\mathrm{m})\)~\cite{Ergun1949}.
Because both relations were originally developed for packed beds of particles rather than fibrous porous media, their application to carbon-felt electrodes should be regarded as approximate.
For parameterization specific to a given porous electrode, \(K\) and \(\beta\) can be determined by fitting the Darcy--Forchheimer relation to experimental pressure-drop data~\cite{Heijden2023redoxexperimental}.

For a rigid porous electrode with uniform porosity \(\varepsilon\), steady incompressible flow, constant electrolyte density \(\rho\) and viscosity \(\mu\), and \(\alpha=1\), \cref{eq:NavierStokes} simplifies to
\begin{subequations}\label{eq:N_S_B_D_F_incomp}
\begin{align}
    \frac{\rho}{\varepsilon^2}
    \left(\bm v\cdot\nabla\right)\bm v
    &=
    -\nabla p
    +
    \frac{\mu}{\varepsilon}\Delta\bm{v}
    -
    \frac{\mu}{K}\bm{v}
    -
    \rho\beta |\bm{v}|\bm{v},
    \label{eq:momentum_nsdf}
    \\
    \nabla\cdot\bm{v}
    &=0 .
    \label{eq:continuity_NSDF}
\end{align}
\end{subequations}
A formulation of this form is also used in RfbFoam~\cite{Barzegari2026RfbFoam}.
In the 3D model, \cref{eq:N_S_B_D_F_incomp} is used to describe the electrolyte flow in both the porous-electrode and flow-channel regions; in the non-porous flow-channel region, it reduces to the incompressible Navier--Stokes equations, corresponding to $\varepsilon=1$, $K^{-1}=0$, and $\beta=0$.
In the finite-element implementation, the porous-electrode and flow-channel contributions are assembled separately over \(\Omega_E\) and \(\Omega_F\), respectively.

Resolving the full volume-averaged flow system~\eqref{eq:N_S_B_D_F_incomp} is computationally demanding and may be unnecessary for models without flow channels.
For the 2D model, we assume that the leading-order momentum balance is between the pressure gradient and Darcy resistance, while the convective inertial, Brinkman viscous diffusion, and Forchheimer resistance terms are negligible. 
Accordingly, \cref{eq:N_S_B_D_F_incomp} reduces to
\begin{subequations}\label{eq:darcy-flow}
\begin{align}
\bm{v} &= -\frac{K}{\mu}\nabla p,
\label{eq:darcy-law}
\\
\nabla\cdot\bm{v} &= 0.
\label{eq:continuity-darcy}
\end{align}
\end{subequations}
Here, \cref{eq:darcy-law} is Darcy's law and
\cref{eq:continuity-darcy} expresses incompressibility.

\subsubsection{Summary of equations for the VRFB negative half-cell}

So far, we have presented the governing equations for a general RFB half-cell.
We now specialize the formulation to the negative half-cell of a VRFB and use the corresponding chemistry hereafter, while treating the \mbox{RfbFoam} verification case separately.

The negative half-cell of a VRFB contains an aqueous sulfuric acid solution with the redox-active species \(\mathrm{V}^{2+}\) and \(\mathrm{V}^{3+}\), corresponding to the reduced and oxidized forms, respectively, together with the supporting ions \(\mathrm{H}^{+}\), \(\mathrm{HSO}_4^{-}\), and \(\mathrm{SO}_4^{2-}\).
In this case, the general half-reaction \eqref{eq:negative_half_cell} gives
\begin{equation}
\mathrm{V}^{3+} + \mathrm{e}^{-}
\xrightleftharpoons[\text{discharge}]{\text{charge}}
\mathrm{V}^{2+},
\label{eq:vrfb_negative_reaction}
\end{equation}
and the set of ionic species included in the model is
$\mathcal{J}=\{\mathrm{V}^{2+},\, \mathrm{V}^{3+},\, \mathrm{H}^{+},\,\mathrm{HSO}_4^{-},\, \mathrm{SO}_4^{2-}\}$, from which we exclude $\mathrm{H_2O}$ as it does not enter our governing equations.
Of these five ionic concentrations, only four are independent; we single out the concentration of \(\mathrm{SO}_4^{2-}\), which we obtain from the local electroneutrality constraint in \cref{eq:electroneutrality},
\begin{align}\label{eq:charge-neutrality-sulfate}
    c_{\mathrm{SO}_4^{2-}} = -\frac{1}{z_{\mathrm{SO}_4^{2-}}} \sum_{j\in\mathcal{J}\setminus\{\mathrm{SO}_4^{2-}\}} z_j c_j.
\end{align}

Equations \eqref{eq:mass_balance}, \eqref{eq:Nernst-Planck},
\eqref{eq:species_source_terms},
\eqref{eq:electrolyte-current-density-compact},
\eqref{eq:solid_ohms_law}--\eqref{eq:Eeq}, and
\eqref{eq:surface_concentration} then reduce to six governing
electrochemical equations,
\begin{widetext}
\begin{subequations}\label{eq:governing-system}
\begin{align}
&\frac{\partial \!\left(\varepsilon\, c_j\right)}{\partial t}
+\nabla\!\cdot \left(
-D_j^{\mathrm{eff}}\nabla c_j
-\frac{z_j c_j D_j^{\mathrm{eff}}}{RT}F\nabla\phi_\ell
+\bm{v}\,c_j
\right)
=
\begin{cases}
-\dfrac{a i_F}{F}, & j=\mathrm{V}^{2+},\\[6pt]
+\dfrac{a i_F}{F}, & j=\mathrm{V}^{3+},\\[6pt]
0, & j\in\{\mathrm{H}^{+},\mathrm{HSO}_4^{-}\},
\end{cases}
\label{eq:VRFB_mass_balance}
\\
&\nabla\cdot\bm{i}_{\ell}
=
\nabla\cdot\left[
-\left(
\frac{F^2}{RT}
\sum_{j\in\mathcal{J}}z_j^2D_j^{\mathrm{eff}}c_j
\right)\nabla\phi_\ell
-F\sum_{j\in\mathcal{J}}z_jD_j^{\mathrm{eff}}\nabla c_j
\right]
=
a\,i_F,
\label{eq:i_ell_VRFB}
\\
&\nabla\cdot\bm{i}_{s}
=
\nabla\cdot\left(
-\sigma^{\mathrm{eff}}\nabla\phi_s
\right)
=
-a\,i_F,
\label{eq:i_s_VRFB}
\\
&i_F
=
Fk^{0}
\left(c_{\mathrm{V}^{2+}}\right)^{\alpha_c}
\left(c_{\mathrm{V}^{3+}}\right)^{\alpha_a}
\left[
\frac{c_{\mathrm{V}^{2+}}^{\mathrm{surf}}}
     {c_{\mathrm{V}^{2+}}}
\exp\!\left(\frac{\alpha_aF}{RT}\eta\right)
-
\frac{c_{\mathrm{V}^{3+}}^{\mathrm{surf}}}
     {c_{\mathrm{V}^{3+}}}
\exp\!\left(-\frac{\alpha_cF}{RT}\eta\right)
\right].
\label{eq:i_F_VRFB_expanded}
\end{align}
\end{subequations}
\end{widetext}
The overpotential and surface concentrations in \cref{eq:i_F_VRFB_expanded} are given by
\cref{eq:overpotential,eq:Eeq,eq:surface_concentration}, with \(\mathcal{R}=\mathrm{V}^{2+}\) and \(\mathcal{O}=\mathrm{V}^{3+}\).
The system of electrochemical relations given in \cref{eq:governing-system}, together with the corresponding electrolyte-flow equations, \cref{eq:N_S_B_D_F_incomp} for the 3D model and \cref{eq:darcy-flow} for the 2D model, forms a closed set of equations for eight primary unknown fields: four ionic concentration fields \(c_j\), the solid- and liquid-phase electric potentials \(\phi_s\) and \(\phi_\ell\), the electrolyte velocity field \(\bm{v}\), and the pressure \(p\).

First, note that the charge-conservation relations in \cref{eq:i_ell_VRFB,eq:i_s_VRFB} are coupled through the Faradaic-current-density expression \eqref{eq:i_F_VRFB_expanded} and form a coupled nonlinear system for the solid- and electrolyte-phase potentials, $\phi_s$ and $\phi_\ell$. 
Because the Faradaic current density also depends on the species concentrations, the potential and concentration fields are coupled as well.
Second, note that the flow equations \eqref{eq:N_S_B_D_F_incomp} and \eqref{eq:darcy-flow} do not contain the electric potentials, Faradaic current density, ionic concentrations, or ionic fluxes. 
Conversely, the velocity field enters the species-transport equation \eqref{eq:VRFB_mass_balance} through the advective flux term.
Hence, in our half-cell model, electrolyte flow is coupled one-way to the electrochemical transport problem, simplifying the numerical routines described in \cref{sec:methods}. 
In a full-cell model, membrane-driven electro-osmotic flow and other mechanisms may couple ionic transport back to the electrolyte flow.

\subsection{Boundary conditions}

\subsubsection{Species transport}
Let \(\mathcal{J}_{\mathrm{ind}}=\mathcal{J}\setminus\{\mathrm{SO}_4^{2-}\}\) denote the set of independently solved ionic species. 
For \(j\in\mathcal{J}_{\mathrm{ind}}\), we use the following species-transport boundary conditions for the 2D and 3D models:
\begin{subequations}\label{eq:bc-species}
\begin{align}
&c_j = c_j^{\mathrm{res}}
&& \text{on } \Gamma_{\mathrm{in}},
\label{eq:bc-species-inlet}
\\
&\left(
-D_j^{\mathrm{eff}}\nabla c_j
-\frac{z_j F D_j^{\mathrm{eff}}}{RT}c_j\nabla\phi_\ell
\right)\!\cdot\!\bm{n} = 0
&& \text{on } \Gamma_{\mathrm{out}},
\label{eq:bc-species-outlet-nonconv}
\\
&\bm{N}_{\mathrm{H}^{+}}\cdot\bm{n}
= \frac{\bm{i}_{\ell}\cdot\bm{n}}{F}
&& \text{on } \Gamma_{\mathrm{mem}},
\label{eq:bc-species-proton-membrane}
\\
&\bm{N}_j\cdot\bm{n} = 0,
\qquad
j\in\mathcal{J}_{\mathrm{ind}}\setminus\{\mathrm{H}^{+}\}
&& \text{on } \Gamma_{\mathrm{mem}},
\label{eq:bc-species-blocked-membrane}
\\
&\bm{N}_j\cdot\bm{n} = 0
&& \text{on } \Gamma_{\mathrm{cc}}\cup\Gamma_{\mathrm{wall}} .
\label{eq:bc-species-cc-wall}
\end{align}
\end{subequations}
First, \cref{eq:bc-species-inlet} fixes the ionic concentrations at the inlet to remain at prescribed constant reservoir values $c_j^{\mathrm{res}}$, corresponding to a half-cell connected to a much larger storage tank. 
Second, \cref{eq:bc-species-outlet-nonconv} follows under the assumption of advection-dominated outflow \cite{SHAH20088087}, \(\bm{N}_j\cdot\bm{n}=(\bm{v}\cdot\bm{n})c_j\).
Next, \cref{eq:bc-species-proton-membrane,eq:bc-species-blocked-membrane} describe an ideal membrane that allows only protons to pass while blocking all other ionic species.
Finally, \cref{eq:bc-species-cc-wall} imposes impermeability to all species at the current collector and wall boundaries.

For the 1D model, constant concentrations are prescribed at the membrane:
\begin{align}
c_j &= c_j^{\mathrm{mem}}
&&\text{on }\Gamma_{\mathrm{mem}},
\\
\bm N_j\cdot\bm n &=0
&&\text{on }\Gamma_{\mathrm{cc}}.
\end{align}

\subsubsection{Current and potential}\label{sec:electricbcs}
In RFBs, electronic current enters or leaves the domain only through the current collector, whereas ionic current does so only through the membrane; hence,
\begin{subequations}\label{eq:bs-currents-both-cases}
\begin{align}
\bm{i}_s\cdot\bm{n} &= 0
&& \text{on } \partial\Omega \setminus \Gamma_{\mathrm{cc}} ,\\
\bm{i}_\ell\cdot\bm{n} &= 0
&& \text{on } \partial\Omega \setminus \Gamma_{\mathrm{mem}}.
\end{align}
\end{subequations}
The remaining boundary conditions at the current collector and membrane depend on whether the RFB is operated under voltage control, current control, or combined current–voltage protocols~\cite{Banguero2018BatteryChargingControl}. 

\paragraph{Potentiostatic mode}\label{sec:potentiostatic}

In potentiostatic mode, the half-cell potential difference, $V_{\mathrm{app}} =\phi_s|_{\Gamma_{\mathrm{cc}}}-\phi_\ell|_{\Gamma_{\mathrm{mem}}}$, between the solid phase at the current collector and the electrolyte phase at the membrane is prescribed.
We choose to ground the solid-phase potential at the current collector, 
\begin{align}\label{eq:bc-pot-solid}
\phi_s &= 0
\qquad
&& \text{on } \Gamma_{\mathrm{cc}}, 
\end{align}
yielding, for the electrolyte potential at the membrane, 
\begin{align}\label{eq:bc-pot-elyte}
\phi_\ell &= -V_{\mathrm{app}}
&& \text{on } \Gamma_{\mathrm{mem}}. 
\end{align}

\paragraph{Galvanostatic mode}\label{sec:galvanostatic}

In galvanostatic mode, the applied current density \(i_{\mathrm{app}}\) is imposed through normal current-flux boundary conditions: for the electronic current in the solid phase,
\begin{align}
\bm{i}_{s}\!\cdot\!\bm{n} &= -i_{\mathrm{app}}
&& \text{on } \Gamma_{\mathrm{cc}}; \label{eq:bc-solid-left}
\end{align}
for the ionic current in the electrolyte phase,
\begin{align}
\bm{i}_{\ell}\!\cdot\!\bm{n} &= i_{\mathrm{app}}
&& \text{on } \Gamma_{\mathrm{mem}}. \label{eq:bc-elyte-right}
\end{align}
Here, positive values of \(i_{\mathrm{app}}\) correspond to current in the positive \(x\)-direction, whereas negative values correspond to current in the negative \(x\)-direction.
A relative minus sign difference appears between \cref{eq:bc-solid-left,eq:bc-elyte-right} because the outward unit normal vector on the boundary, \(\bm{n}\), points in the negative \(x\)-direction on \(\Gamma_{\mathrm{cc}}\) and in the positive \(x\)-direction on \(\Gamma_{\mathrm{mem}}\). 

As the electrochemical system \eqref{eq:governing-system} is invariant under $\phi_s \mapsto \phi_s + C$ and $\phi_\ell \mapsto \phi_\ell + C$, with $C\in\mathbb{R}$, the solid and electrolyte potentials $\phi_s$ and $\phi_\ell$ are defined up to an additive constant---this constant drops in the overpotential $\eta$, so $\eta$ and the resulting currents are uniquely defined.
In potentiostatic mode, we fix the constant $C$ through \cref{eq:bc-pot-solid}; in galvanostatic mode, the constant is not specified.
This gauge freedom in galvanostatic mode leads to singular behavior discussed in \cref{sec:implementation}, which, from a numerical perspective, makes galvanostatic mode more challenging than potentiostatic mode.
Reference-potential and null-space treatments for this type of galvanostatic current-flux porous-electrode problem are discussed by Wang et al.~\cite{Wang2025_NonlinearlyCoupledPoisson}.

\subsubsection{Electrolyte flow}
For the 3D electrolyte-flow problem, we use the boundary conditions 
\begin{subequations}\label{eq:bc-flow}
\begin{align}
\bm{v} &= \bm{v}_{\mathrm{in}}
&&\text{on } \Gamma_{\mathrm{in}},
\label{eq:bc-flow-inlet}
\\
\left(-p\bm I+\mu\nabla\bm v\right)\bm n
&=
-p_{\mathrm{out}}\bm n
&&\text{on } \Gamma_{\mathrm{out}},
\label{eq:bc-flow-outlet-traction}
\\
\bm{v} &= \bm{0}
&& \text{on }
\partial\Omega \setminus
\left(\Gamma_{\mathrm{in}}\cup\Gamma_{\mathrm{out}}\right),
\label{eq:bc-flow-noslip}
\end{align}
\end{subequations}
where \cref{eq:bc-flow-inlet} imposes the velocity \(\bm v_{\mathrm{in}}\) at the inlet, \cref{eq:bc-flow-outlet-traction} relates the pressure and viscous stress at the outlet to the reference pressure \(p_{\mathrm{out}}\), and \cref{eq:bc-flow-noslip} imposes no-slip conditions on the remaining boundaries.

For the 2D Darcy flow problem, we use 
\begin{subequations}\label{eq:bc-darcy}
\begin{align}
-\frac{K}{\mu}\nabla p\cdot\bm n
&= -v_{\mathrm{in}}
&& \text{on } \Gamma_{\mathrm{in}},
\label{eq:bc-darcy-inlet}
\\
p &= p_{\mathrm{out}}
&& \text{on } \Gamma_{\mathrm{out}},
\label{eq:bc-darcy-outlet}
\\
-\frac{K}{\mu}\nabla p\cdot\bm n
&= 0
&& \text{on }
\partial\Omega \setminus
\left(\Gamma_{\mathrm{in}}\cup\Gamma_{\mathrm{out}}\right),
\label{eq:bc-darcy-walls}
\end{align}
\end{subequations}
where \cref{eq:bc-darcy-inlet} relates the pressure gradient at the inlet to the prescribed inlet velocity $v_{\mathrm{in}}$, \cref{eq:bc-darcy-outlet} sets the pressure at the outlet to a prescribed value $p_{\mathrm{out}}$, and \cref{eq:bc-darcy-walls} signifies impermeability of the remaining boundaries.

\subsection{Initial conditions}

At \(t=0\), we set the independently solved species concentrations to their reservoir values:
\begin{equation}
c_j(\bm{x},0)=c_j^{\mathrm{res}},
\qquad
\bm{x}\in\Omega,
\qquad
j\in\mathcal{J}\setminus
\left\{\mathrm{SO}_4^{2-}\right\}.
\label{eq:representative_initial_conditions}
\end{equation}
The initial sulfate concentration was then calculated from the electroneutrality condition.

\section{Methods}\label{sec:methods}

\subsection{Numerical implementation}\label{sec:implementation}

The 1D and 2D models use \(N_x\)-element and structured \(N_x \times N_y\) meshes, respectively, while the 3D model uses an externally generated mesh with cell and facet tags to identify the subdomains and boundary regions defined in \cref{sec:model}.

Under galvanostatic operation, \software{} does not introduce an additional reference-potential constraint to fix the potentials in the solid and electrolyte phases. 
These potentials therefore retain the common additive gauge freedom discussed above for galvanostatic operation.
Instead, the linearized systems arising in the Newton iterations are solved using the sparse direct solver MUMPS~\cite{Amestoy2001MUMPS} through PETSc~\cite{balay2019petsc}. 
For compatible singular systems, this allows a representative solution of the galvanostatic current-flux formulation to be obtained without modifying the discrete equations or adding a problem-specific gauge condition. 
The resulting absolute values of \(\phi_s\) and \(\phi_\ell\) depend on the arbitrary gauge, whereas gauge-invariant quantities such as potential gradients and overpotentials are unaffected.

\subsection{Weak formulation and time discretization}

For the 2D and 3D models, the flow problem is solved first, using either the Navier--Stokes formulation with Darcy--Forchheimer resistance~\eqref{eq:N_S_B_D_F_incomp} or the Darcy formulation~\eqref{eq:darcy-flow}, and the resulting velocity field is then treated as a known input in the coupled species- and charge-transport problem governed by the electrochemical system \eqref{eq:governing-system}.

In the 3D model, the velocity and pressure fields are computed from \cref{eq:N_S_B_D_F_incomp} by solving a coupled velocity--pressure flow problem on the combined porous-electrode and flow-channel domain. 
Let \(U_h\) and \(Q_h\) denote the finite-element spaces for the velocity and pressure fields, respectively. 
We use the inf-sup stable Taylor--Hood element pair, with continuous quadratic Lagrange elements for the velocity and continuous linear Lagrange elements for the pressure,
\[
U_h=[\mathbb{P}_2]^{n_{\mathrm{dim}}},\qquad Q_h=\mathbb{P}_1 .
\]
The corresponding velocity and pressure test functions are denoted by \(\bm w_h\in U_h\) and \(q_h\in Q_h\), respectively.
The weak form is: find \((\bm v_h,p_h)\in U_h\times Q_h\) such that, for all \((\bm w_h,q_h)\in U_h\times Q_h\),
\begin{equation}
\begin{aligned}
&\int_{\Omega_E}
\frac{\rho}{\varepsilon^2}
\left(\bm v_h\cdot\nabla\right)\bm v_h\cdot\bm w_h\,dx
+
\int_{\Omega_F}
\rho
\left(\bm v_h\cdot\nabla\right)\bm v_h\cdot\bm w_h\,dx\\
&-
\int_{\Omega_E\cup\Omega_F}
p_h\nabla\cdot\bm w_h\,dx
+
\int_{\Omega_E}
\frac{\mu}{\varepsilon}
\nabla\bm v_h:\nabla\bm w_h\,dx
\\
&+
\int_{\Omega_F}
\mu
\nabla\bm v_h:\nabla\bm w_h\,dx
+
\int_{\Omega_E}
\frac{\mu}{K}\bm v_h\cdot\bm w_h\,dx
\\
&+
\int_{\Omega_E}
\rho\beta |\bm v_h|\bm v_h\cdot\bm w_h\,dx
-
\int_{\Omega_E\cup\Omega_F}
q_h\nabla\cdot\bm v_h\,dx
\\
&+
\int_{\Gamma_{\mathrm{out}}}
p_{\mathrm{out}}\,\bm n\cdot\bm w_h\,ds
=0 .
\label{eq:app-flow-final-weak-form_txt}
\end{aligned}
\end{equation}
The Darcy and Forchheimer resistance terms are applied only in the porous-electrode region \(\Omega_E\); they are omitted in the non-porous flow-channel region \(\Omega_F\). 
The derivation leading to \cref{eq:app-flow-final-weak-form_txt} is presented in Appendix~\ref{app:bulk-flow-weak-form}.

In the 2D porous-electrode model, the Darcy formulation~\eqref{eq:darcy-flow} is used to describe the fluid flow. 
Let \(Q_h^{\mathrm D}\) denote the finite-element space for the Darcy pressure field.
Thus, our task is to find \(p_h\in Q_h^{\mathrm D}\), satisfying the prescribed outlet-pressure condition, such that
\begin{equation}
\begin{aligned}
&\int_{\partial\Omega}
\left(-\frac{K}{\mu}\nabla p_h\cdot\bm n\right)q_h\,ds
+
\int_{\Omega}
\frac{K}{\mu}\nabla p_h\cdot\nabla q_h\,dx
=0,
\\
&\qquad\qquad\qquad
\forall q_h\in Q_{h,0}^{\mathrm D},
\end{aligned}
\label{eq:weak_darcy_pressure}
\end{equation}
where \(Q_{h,0}^{\mathrm D}\) denotes the corresponding pressure test space with homogeneous outlet-pressure condition.

After obtaining the flow field, we solve the electrochemical system \eqref{eq:governing-system}, treating the velocity field as a known input.
For the electrochemical problem, conforming Lagrange finite elements are used for the species concentrations and electric potentials. 
Let \(V=H^1(\Omega)\) and let \(V_h\subset V\) denote the continuous Lagrange finite-element space of the chosen polynomial degree. 
The primary electrochemical unknowns are collected in the mixed finite-element space $
\mathbb{V}_h = V_h^6$. 
The discrete electrochemical solution space is written as
\[
X_h =
(\phi_{s,h},\phi_{\ell,h},
c_{\mathrm{V}^{2+},h},
c_{\mathrm{V}^{3+},h},
c_{\mathrm{H}^{+},h},
c_{\mathrm{HSO}_4^{-},h})
\in\mathbb{V}_h,
\]
with corresponding test space
\[
Y_h =
(v_{s,h},v_{\ell,h},
v_{\mathrm{V}^{2+},h},
v_{\mathrm{V}^{3+},h},
v_{\mathrm{H}^{+},h},
v_{\mathrm{HSO}_4^{-},h})
\in\mathbb{V}_h .
\]
The sulfate concentration is recovered algebraically by \cref{eq:charge-neutrality-sulfate}. 
The electrolyte-phase charge-conservation equation~\eqref{eq:i_ell_VRFB} is written in weak form as
\begin{equation}
\int_{\partial\Omega}
\bm{i}_{\ell}\cdot\bm n\,v_{\ell,h}\,ds
-
\int_{\Omega}
\bm{i}_{\ell}\cdot\nabla v_{\ell,h}\,dx
-
\int_{\Omega_E}
a i_F v_{\ell,h}\,dx
=0 .
\end{equation}
The solid-phase charge-conservation equation~\eqref{eq:i_s_VRFB} is written as
\begin{equation}
\int_{\partial\Omega_E}
\bm i_s\cdot\bm n\,v_{s,h}\,ds
-
\int_{\Omega_E}
\bm i_s\cdot\nabla v_{s,h}\,dx
+
\int_{\Omega_E}
a i_F v_{s,h}\,dx
=0 .
\end{equation}
The boundary terms are evaluated using the galvanostatic or potentiostatic boundary conditions described in \cref{sec:electricbcs}.

For transient simulations, the species balances are discretized using a first-order implicit scheme. 
For each independent species, the weak form at time level \(n+1\) is
\begin{equation}
\begin{aligned}
\int_{\Omega}
\frac{\varepsilon}{\Delta t}
\left(c_{j,h}^{n+1}-c_{j,h}^{n}\right)
v_{j,h}\,dx
+
\int_{\partial\Omega}
\bm N_j^{n+1}\cdot\bm n\,v_{j,h}\,ds
\\
-
\int_{\Omega}
\bm N_j^{n+1}\cdot\nabla v_{j,h}\,dx
-
\int_{\Omega_E}
S_j^{n+1}v_{j,h}\,dx =0.
\end{aligned}
\end{equation}
The electric potentials are treated as quasi-steady at each time level, whereas the time-derivative term is retained only in the species balances.
For steady-state simulations, the first term in the species weak form is omitted.

During software development, ChatGPT (OpenAI) was used to assist with code refactoring and to improve code readability. The authors reviewed and tested all AI-assisted code modifications.  

\section{Results and discussion}\label{sec:results}

We first verify \software{} by comparing its predictions with independent implementations and published benchmark solutions. These comparison cases correspond to limiting parameter regimes of the general formulation implemented in \software{}.
After the verification, we show representative transient simulations to illustrate the physical insight that \software{} can provide.

\subsection{Uniform-concentration charge-conservation limit}

\subsubsection{Formulation of RFB at high electrolyte flow and galvanostatic mode}

Under galvanostatic operation, increasing the inlet flow rate reduces reaction-driven concentration changes by replenishing the reactants and carrying reaction products out of the cell.
Thus, in the limit of sufficiently high electrolyte flow rate, the ionic concentrations can be approximated as time-independent.
More precisely, this regime is reached for large Péclet and small Damköhler numbers, as we show in Appendix~\ref{app:specific-regime}.
A similar approximation is also valid immediately after the circuit is closed at $t=0^+$, when the electrolyte still reflects its initially uniform composition~\cite{Newman1962PorousCurrentDistribution}.
We refer to this approximation as the uniform-concentration charge-conservation (UCC) limit. 

In this regime, since the species concentrations equal their inlet concentrations throughout the cell, we can drop the species mass conservation equations. Additionally, because
\(\nabla c_j=0\), we can drop the diffusion contribution to the ionic current. So from the governing equations \eqref{eq:governing-system},
we no longer need \cref{eq:VRFB_mass_balance}, and \cref{eq:i_ell_VRFB} simplifies to
\begin{equation}\label{eq:ucc_phi_l}
    \nabla\!\cdot\!\big(-\kappa^{\mathrm{eff}} \nabla \phi_\ell\big)
= a\, i_F ,
\end{equation}
where \(\kappa^{\mathrm{eff}}\) is evaluated using the prescribed inlet concentrations. 
The fixed concentrations are also used to evaluate the equilibrium potential and exchange current density, which enter the Butler--Volmer expression for the Faradaic current density.
Thus, the reduced UCC formulation solves only the solid- and electrolyte-phase charge-conservation equations for
\(\phi_s\) and \(\phi_\ell\).

\begin{figure}
  \centering
    \subfloat[]{
        \includegraphics[width=\linewidth]
        {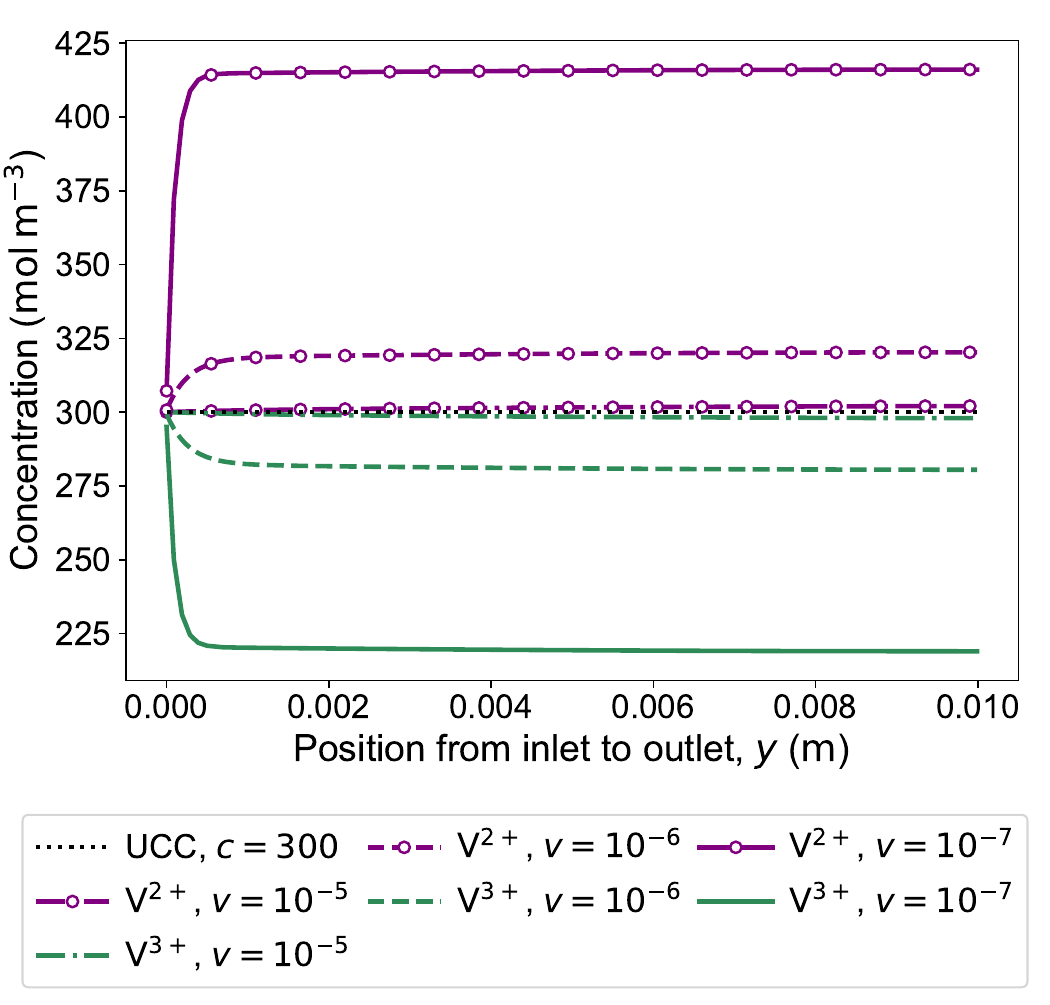}
    }
    \par\medskip
    
    \subfloat[]{
        \includegraphics[width=\linewidth]
        {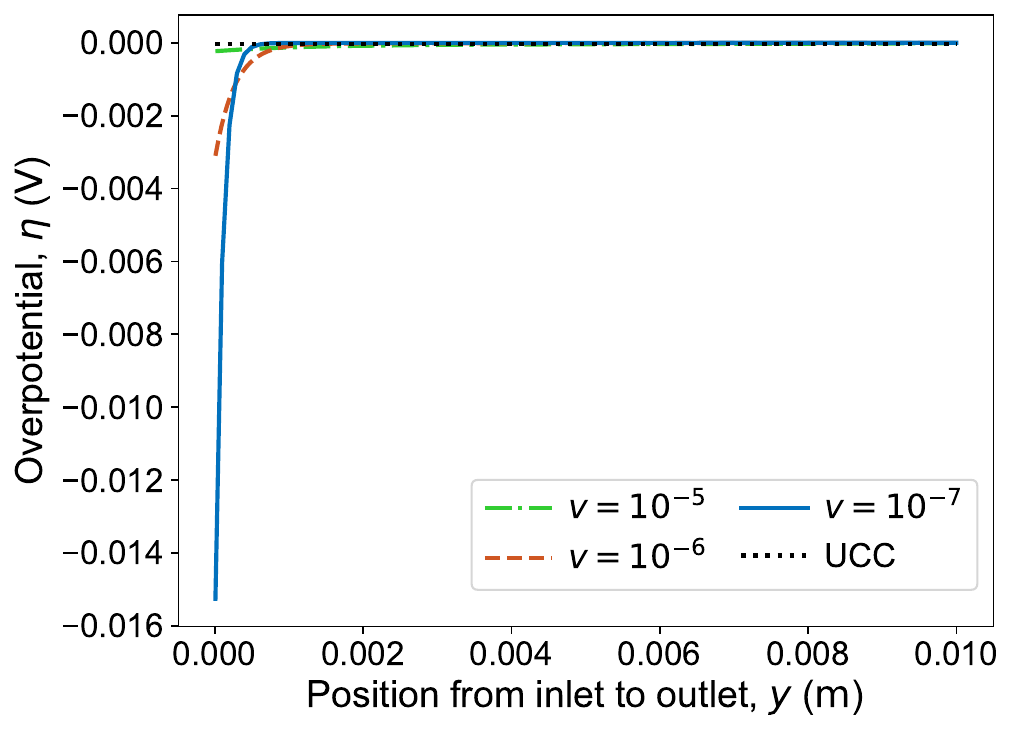}
    }
    
  \caption{\software{} results comparing the full 2D model at different inlet velocities with the reduced formulation. (a) Redox-active-species concentrations along the flow direction at \(x=W/2\); the UCC line denotes the prescribed inlet concentration. (b) Corresponding overpotential profiles.}

  \label{fig:cc_limit_profiles}
\end{figure}

To study how electrolyte flow affects the approach to the UCC limit,  we used the 2D version of \software{} with the parameters in \cref{tab:representative_physical_params} and performed simulations for different inlet velocities, $\bm{v}_{\mathrm{in}}$, while keeping all other parameters fixed.
In this table, several material and electrochemical parameters were adopted from or derived using data reported by Muñoz-Perales et al.~\cite{MunozPerales2022OperationVariablesVRFB}.
\Cref{fig:cc_limit_profiles} shows the concentration profiles of the redox-active species \(c_{\mathrm{V}^{2+}}\) and \(c_{\mathrm{V}^{3+}}\) in panel~(a) and the overpotential profile in panel~(b), along the \(y\)-direction at fixed \(x=W/2\).
To generate \cref{fig:cc_limit_profiles}, we used two formulations.
First, we solved the full 2D formulation, comprising the system of electrochemical equations \eqref{eq:governing-system} coupled to the Darcy equation \eqref{eq:darcy-flow}, for several inlet velocities.
Second, we solved the reduced UCC formulation described around \cref{eq:ucc_phi_l}, in which the concentrations are
prescribed at their inlet values, and only the two charge-conservation equations are solved. 
We see that, for low inlet velocity, $c_{\mathrm{V}^{2+}}>c_{\mathrm{V}^{3+}}$, meaning that the reaction product $\mathrm{V}^{2+}$ remains in the cell for some time.
As the inlet velocity increases, the difference in concentration becomes progressively smaller; at $v_{\mathrm{in}}=10^{-5}\,\mathrm{m\,s^{-1}}$, the concentrations halfway along the flow path are roughly equal to their inlet values.
The corresponding overpotential profiles show the same trend. Increasing $v_{\mathrm{in}}$ reduces the spatial variation of the overpotential; at $v_{\mathrm{in}}=10^{-5}\,\mathrm{m\,s^{-1}}$, the profile is nearly uniform and close to the UCC result.

\begin{table}[t]
\centering
\caption{Geometric, material, kinetic, and operating parameters used for the VRFB half-cell simulations.}
\label{tab:representative_physical_params}
\scriptsize
\setlength{\tabcolsep}{4pt}
\renewcommand{\arraystretch}{1.15}
\begin{tabular}{@{}l l@{}}
\toprule
Quantity & Value (unit) \\
\midrule

\multicolumn{2}{@{}l@{}}{\textit{Computational geometry}} \\

Model dimensionality
& \(2\mathrm{D}\) \\

Porous-electrode width, \(W\)
& \(2.5\times10^{-3}\ \mathrm{m}\)\\

Porous-electrode length, \(L\)
& \(1.0\times10^{-2}\ \mathrm{m}\) \\

\midrule
\multicolumn{2}{@{}l@{}}{\textit{Porous-electrode and electrolyte properties}} \\

Intrinsic solid conductivity, \(\sigma\)
& \(3402.1\ \mathrm{S\,m^{-1}}\) \\

Specific interfacial area, \(a\)
& \(2.3\times10^{4}\ \mathrm{m^{-1}}\) \\

Porous-electrode porosity, \(\varepsilon\)
& \(0.94\) \\

Electrolyte dynamic viscosity, \(\mu\)
& \(6.963\times10^{-3}\ \mathrm{Pa\,s}\) \\

Electrolyte density, \(\rho\)
& \(1399.8\ \mathrm{kg\,m^{-3}}\) \\

Kozeny--Carman constant, \(C_{\mathrm{KC}}\)
& \(5.55\) \\

Carbon-fiber diameter, \(d_f\)
& \(0.5\times10^{-6}\ \mathrm{m}\) \\

\midrule
\multicolumn{2}{@{}l@{}}{\textit{Electrochemical and operating parameters}} \\
Charge-transfer coefficients, \(\alpha_a=\alpha_c\)
& \(0.5\) \\

Standard rate constant, \(k^0\)
& \(2.5\times10^{-7}\ \mathrm{m\,s^{-1}}\) \\

Formal potential, \(E_0'\)
& \(-0.36\ \mathrm{V}\) \\

Temperature, \(T\)
& \(298.15\ \mathrm{K}\) \\

Applied current-density magnitude, \(\abs{i_{\mathrm{app}}}\)
& \(0.5\ \mathrm{A\,m^{-2}}\) \\

Inlet velocity, \(v_{\mathrm{in}}\)
& \(1.0\times10^{-7}\ \mathrm{m\,s^{-1}}\) \\

\midrule
\multicolumn{2}{@{}l@{}}{\textit{Species concentrations and diffusivities}} \\

Inlet concentration of \(\mathrm{V}^{2+}\),
\(c_{\mathrm{V}^{2+}}^{\mathrm{res}}\)
& \(300.0\ \mathrm{mol\,m^{-3}}\) \\

Inlet concentration of \(\mathrm{V}^{3+}\),
\(c_{\mathrm{V}^{3+}}^{\mathrm{res}}\)
& \(300.0\ \mathrm{mol\,m^{-3}}\) \\

Inlet concentration of \(\mathrm{H}^{+}\),
\(c_{\mathrm{H}^{+}}^{\mathrm{res}}\)
& \(1000.0\ \mathrm{mol\,m^{-3}}\) \\

Inlet concentration of \(\mathrm{HSO}_4^{-}\),
\(c_{\mathrm{HSO}_4^{-}}^{\mathrm{res}}\)
& \(1000.0\ \mathrm{mol\,m^{-3}}\) \\

Diffusivity of \(\mathrm{V}^{2+}\),
\(D_{\mathrm{V}^{2+}}\)
& \(8.10\times10^{-12}\ \mathrm{m^2\,s^{-1}}\) \\

Diffusivity of \(\mathrm{V}^{3+}\),
\(D_{\mathrm{V}^{3+}}\)
& \(1.65\times10^{-11}\ \mathrm{m^2\,s^{-1}}\) \\

Diffusivity of \(\mathrm{H}^{+}\),
\(D_{\mathrm{H}^{+}}\)
& \(9.312\times10^{-9}\ \mathrm{m^2\,s^{-1}}\) \\

Diffusivity of \(\mathrm{HSO}_4^{-}\),
\(D_{\mathrm{HSO}_4^{-}}\)
& \(1.33\times10^{-9}\ \mathrm{m^2\,s^{-1}}\) \\

Diffusivity of \(\mathrm{SO}_4^{2-}\),
\(D_{\mathrm{SO}_4^{2-}}\)
& \(1.065\times10^{-9}\ \mathrm{m^2\,s^{-1}}\) \\
\bottomrule
\end{tabular}
\end{table}

\subsubsection{Verification and performance comparison in UCC limit}

As explained in the preceding subsection, under the UCC assumptions, the full \software{} formulation reduces to coupled solid- and electrolyte-phase charge conservation, closed by the Butler--Volmer relation.
Wang et al.~\cite{Wang2025_NonlinearlyCoupledPoisson} solved the same reduced system and made their numerical implementation publicly available, so we used their implementation for verification.
For a direct comparison, we configured the 2D version of \software{} to reproduce the modeling assumptions of Wang et al.
In the high-flow-rate limit, the mass-transfer coefficients $k_j$ become large, such that the surface and bulk concentrations are approximately equal, $c_\mathcal{R}^{\mathrm{surf}} \approx c_\mathcal{R}$ and $c_\mathcal{O}^{\mathrm{surf}} \approx c_\mathcal{O}$.
Together with the symmetric charge-transfer coefficients, \(\alpha_a=\alpha_c=\tfrac{1}{2}\), this reduces the Butler--Volmer relation to a hyperbolic-sine form.
The resulting equations are
\begin{subequations}\label{eq:model_wang}
\begin{align}
&\nabla\!\cdot\!\big(-\kappa^{\mathrm{eff}} \nabla \phi_\ell\big)
= a\, i_F,
\label{eq:ohms_law_electrolyte}\\
&\nabla\!\cdot\!\big(-\sigma^{\mathrm{eff}} \nabla \phi_s\big)
= -a\, i_F,\label{eq:ohms_law_solid}
\\
  &i_F
  =
  2 i_0 \sinh\left(\frac{F \eta}{2RT}\right).
  \label{eq:BV_sinh}
\end{align}
\end{subequations}
We adopted the geometry, parameter values, and boundary conditions reported by Wang et al., as listed in \cref{tab:Wang}.

\begin{table}
\centering
\caption{Parameters adapted from Ref.~\cite{Wang2025_NonlinearlyCoupledPoisson}.}
\label{tab:Wang}
\scriptsize
\setlength{\tabcolsep}{4pt}
\renewcommand{\arraystretch}{1.15}
\begin{tabular}{@{}l l@{}}
\toprule
Quantity & Value (unit) \\
\midrule
Porous-electrode width, \(W\)
& \(5\times10^{-3}\ \mathrm{m}\) \\

Porous-electrode length, \(L\)
& \(1\times10^{-1}\ \mathrm{m}\) \\

Solid conductivity, \(\sigma\)
& \(1000.0\ \mathrm{S\,m^{-1}}\) \\

Specific interfacial area, \(a\)
& \(1.64\times10^{4}\ \mathrm{m^{-1}}\) \\

Porous-electrode porosity, \(\varepsilon\)
& \(0.78\) \\

Effective electrolyte conductivity, \(\kappa^{\mathrm{eff}}\)
& \(5.9514\ \mathrm{S\,m^{-1}}\) \\

Charge-transfer coefficients, \(\alpha_a=\alpha_c\)
& \(0.5\) \\

Exchange current density, \(i_0\)
& \(2.7657\ \mathrm{A\,m^{-2}}\) \\

Formal potential, \(E_0'\)
& \(-0.1609\ \mathrm{V}\) \\

Temperature, \(T\)
& \(298.15\ \mathrm{K}\) \\

Applied current-density magnitude, \(\abs{i_{\mathrm{app}}}\)
& \(400.0\ \mathrm{A\,m^{-2}}\) \\
\bottomrule
\end{tabular}
\end{table}

For the geometry, parameters, and boundary conditions considered by Wang et al., which are invariant in the \(y\)-direction, the solution is independent of \(y\) and varies only in the through-plane direction \(x\).
Hence, the system of equations~\eqref{eq:model_wang} simplifies to an ordinary differential equation (ODE) for the overpotential along the $x$ coordinate~\cite{Wang2025_NonlinearlyCoupledPoisson}:
\begin{equation}
\frac{\partial^{2}\eta}{\partial x^{2}}
=
2 a i_{0}
\left(
\frac{1}{\sigma^{\mathrm{eff}}}
+
\frac{1}{\kappa^{\mathrm{eff}}}
\right)
\sinh\!\left(\frac{F\eta}{2RT}\right).
\label{eq:eta-ode}
\end{equation}
Importantly, \cref{eq:eta-ode} is a closed equation for the gauge-invariant overpotential $\eta$. 
It therefore avoids the common additive gauge freedom associated with solving \cref{eq:model_wang} for $\phi_{\mathrm{s}}$ and $\phi_{\ell}$ under galvanostatic operation.
We computed a highly resolved solution to the 1D two-point boundary-value problem defined by \cref{eq:eta-ode} using
\texttt{scipy.integrate.solve\_bvp} and used it as a benchmark to assess the 2D implementations of Wang et al. and \software{}.

\begin{figure}
  \centering
    \subfloat[]{
        \includegraphics[width=\linewidth]
        {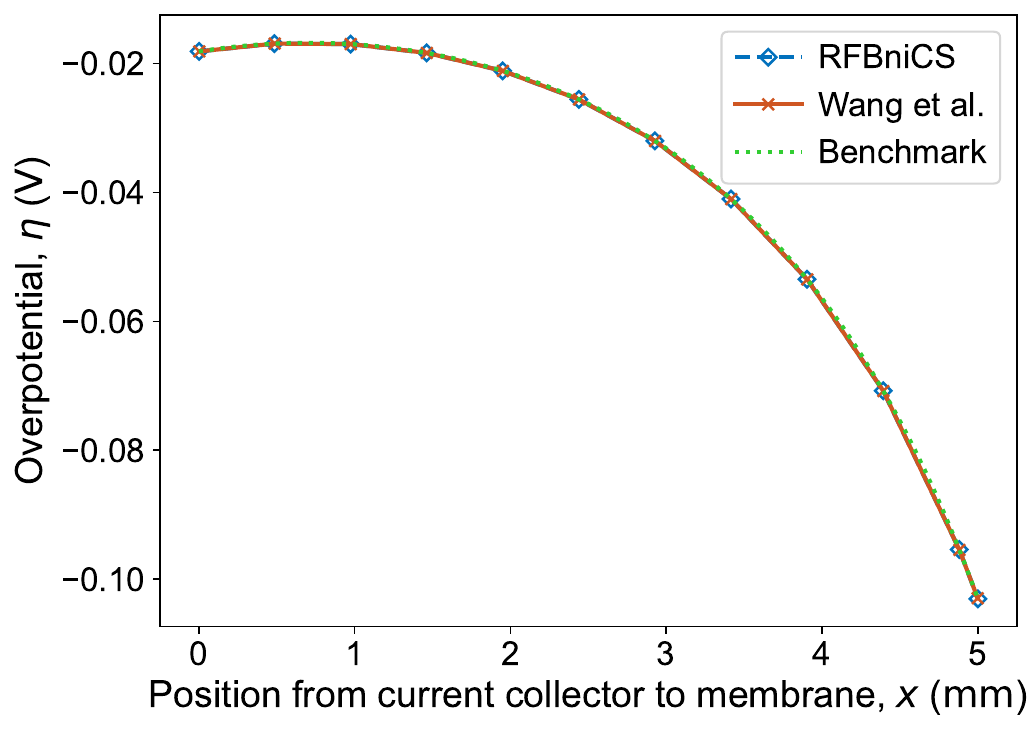}
    }
    \par\medskip
    
    \subfloat[]{
        \includegraphics[width=\linewidth]
        {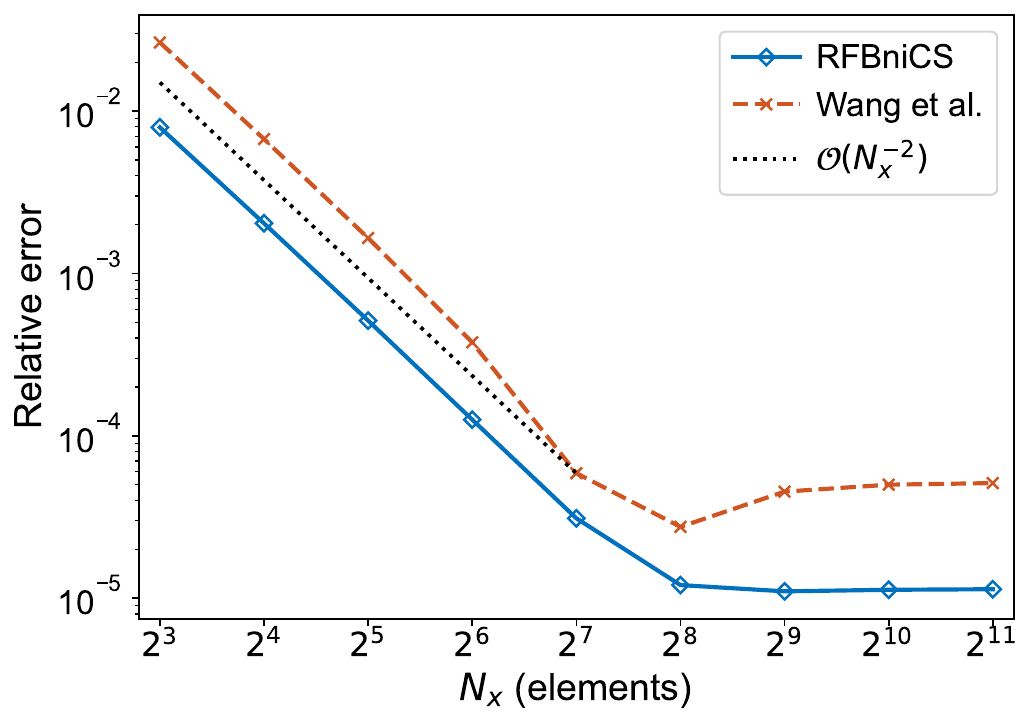}
    }
    \par\medskip
    
    \subfloat[]{
        \includegraphics[width=\linewidth]
        {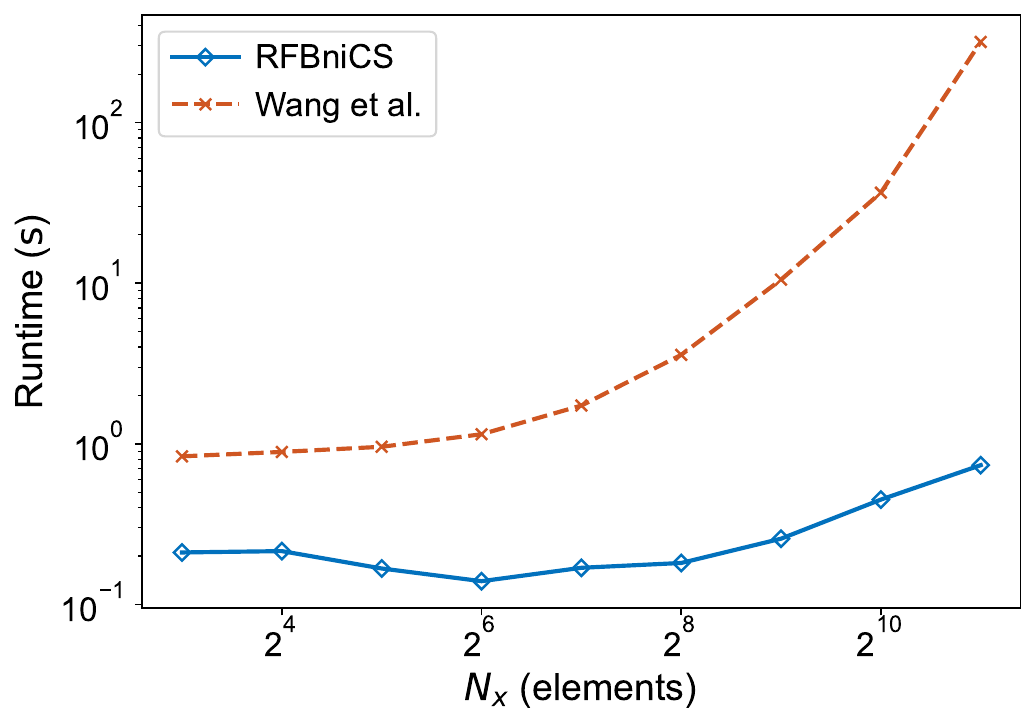}
    }

    \caption{Comparison of \software{} and Wang et al. implementations of \cref{eq:model_wang}, both evaluated along the midline \(y=L/2\), with the 1D benchmark solution from \cref{eq:eta-ode}. (a) Overpotential profile. (b) Relative \(L^2\) error; the dotted line shows the \(\mathcal{O}(N_x^{-2})\) reference slope. (c) Wall-clock runtime.} 

  \label{fig:constant_concentration_limit}
\end{figure}

\Cref{fig:constant_concentration_limit}(a) compares the overpotential profiles from the three models. The 2D profiles, evaluated along the midline $y=L/2$, closely match the benchmark.
In this case of fixed concentrations and kinetic parameters,
the spatial variation of \(\eta\) is only governed by the redistribution of current between the solid and electrolyte phases. 
The larger magnitude of \(\eta\) near the membrane indicates that the Faradaic current density is concentrated near the membrane side of the porous electrode.
This behavior can be explained by the difference between the effective electronic and ionic conductivities.

To quantify the agreement between two scalar fields \(q_a\) and \(q_b\), we define the relative \(L^2\) error over a comparison domain \(\mathcal D\) as
\begin{equation}
\mathcal{E}_{L^2}(q_a,q_b)
=
\frac{\norm{q_a-q_b}_{L^2(\mathcal D)}}
     {\norm{q_b}_{L^2(\mathcal D)}}
=
\frac{
\left(\displaystyle\int_{\mathcal D}
\abs{q_a-q_b}^{2}\,d\mathcal D\right)^{1/2}
}{
\left(\displaystyle\int_{\mathcal D}
\abs{q_b}^{2}\,d\mathcal D\right)^{1/2}
}.
\label{eq:relative-L2-error}
\end{equation}
For the present comparison between the 2D implementations and the 1D benchmark, \(\mathcal D=(0,W)\), with \(q_a(x)=\eta_h(x,L/2)\) and \(q_b(x)=\eta_{\mathrm{ref}}(x)\). 
Using the norm, we investigated the effect of spatial discretization by decreasing the mesh spacing in the \(x\)-direction while keeping the transverse resolution fixed. 
Comparable stopping tolerances were used for both implementations. 
As shown in \cref{fig:constant_concentration_limit}(b), both implementations exhibit approximately second-order convergence over the initial mesh-refinement range, as indicated by the numerical curves being nearly parallel to the second-order reference line $\mathcal{O}(N_x^{-2})$.
The figure also shows that, for the same through-plane resolution, \software{} gives smaller relative errors than the implementation of Wang et al.

\Cref{fig:constant_concentration_limit}(c) compares the corresponding wall-clock runtimes for the 2D models. 
All timings were obtained using the same hardware and the same timing definition. 
This shows that \software{} is faster across all cases, with the performance gap widening with mesh refinement.
This behavior is consistent with the use of optimized finite-element assembly and high-performance nonlinear solver backends in \software{}.

\subsection{Excess supporting electrolyte limit}\label{sec:ese}

\subsubsection{Formulation of RFB in ESE limit}

As the supporting electrolyte concentration, \(C_{\mathcal S}\), increases relative to the redox-active-species concentration, \(C_{\mathcal A}\), the ionic conductivity increases and the electric-potential gradient required to carry the ionic current decreases~\cite{HAVERKORT2024}. 
In the excess-supporting-electrolyte (ESE) limit, \(C_{\mathcal{S}}\gg C_{\mathcal{A}}\),  electromigration of the redox-active species may therefore be neglected~\cite{Barzegari2026RfbFoam,FullerHarb2018}. 
If the concentration-gradient contribution to the electrolyte current density is also negligible, the current density may be approximated by an Ohmic relation~\cite{Barzegari2026RfbFoam,MoraesKelly2023SupportingElectrolyte}.
For the VRFB half-cell in the ESE limit and at steady state, the governing equations \eqref{eq:VRFB_mass_balance} and \eqref{eq:i_ell_VRFB} then reduce to
\begin{subequations}\label{eq:ESE_model}
\begin{align}
\nabla\!\cdot
\left(
-D_j^{\mathrm{eff}}\nabla c_j
+\bm{v}\,c_j
\right)
&=
\begin{cases}
-\dfrac{a i_F}{F},
& j=\mathrm{V}^{2+},\\[6pt]
+\dfrac{a i_F}{F},
& j=\mathrm{V}^{3+},
\end{cases}
\label{eq:ese-species-balances}
\\[6pt]
\nabla\!\cdot\bm{i}_{\ell}
&=
\nabla\!\cdot
\left(
-\kappa_{\mathrm{ESE}}^{\mathrm{eff}}
\nabla\phi_\ell
\right)
=
a i_F,
\label{eq:ese-liquid-charge}
\end{align}
\end{subequations}
while \cref{eq:i_s_VRFB,eq:i_F_VRFB_expanded} and the boundary conditions remain unchanged.
The constant effective ionic conductivity is evaluated from the prescribed reservoir concentrations as
\begin{equation}
\kappa_{\mathrm{ESE}}^{\mathrm{eff}}
=
\frac{F^2}{RT}
\sum_{j\in\mathcal{J}}
z_j^2 D_j^{\mathrm{eff}} c_j^{\mathrm{res}} .
\label{eq:ese-constant-conductivity}
\end{equation}
In the ESE limit, the supporting ions make the dominant contribution to \(\kappa_{\mathrm{ESE}}^{\mathrm{eff}}\), and since their concentrations vary only slightly, the effective ionic conductivity can be assumed constant.
The supporting-electrolyte concentrations, governed by the mass balance in \cref{eq:VRFB_mass_balance}, then decouple from the remaining equations and do not need to be solved. 
Instead, the prescribed reservoir concentrations are used to evaluate the constant effective ionic conductivity.

To investigate how increasing supporting-ion concentrations drives the full formulation toward the ESE limit, we performed steady-state simulations using the 2D version of \software{} with the parameters listed in \cref{tab:representative_physical_params}, except for the varied supporting-ion concentrations and applied current density.
For each electrolyte composition and applied current density, we solved both the full and reduced formulations.
The full formulation comprised the steady-state electrochemical system \eqref{eq:governing-system} coupled to the Darcy-flow equations \eqref{eq:darcy-flow}. 
The reduced formulation used the same Darcy-flow equations and boundary conditions, but replaced \cref{eq:VRFB_mass_balance,eq:i_ell_VRFB} with \cref{eq:ese-species-balances,eq:ese-liquid-charge}, while retaining \cref{eq:i_s_VRFB,eq:i_F_VRFB_expanded}. 
We characterize the electrolyte composition using 
\begin{equation}
C_{\mathcal{S}}^{\mathrm{res}}
=
c_{\mathrm{H}^{+}}^{\mathrm{res}}
+
c_{\mathrm{HSO}_4^{-}}^{\mathrm{res}}
+
c_{\mathrm{SO}_4^{2-}}^{\mathrm{res}},
\qquad
C_{\mathcal{A}}^{\mathrm{res}}
=
c_{\mathrm{V}^{2+}}^{\mathrm{res}}
+
c_{\mathrm{V}^{3+}}^{\mathrm{res}} ,
\end{equation}
where \(C_{\mathcal{S}}^{\mathrm{res}}\) is the total supporting-electrolyte concentration prescribed at the inlet, and \(C_{\mathcal{A}}^{\mathrm{res}}\) is the corresponding total inlet concentration of the vanadium species.
Simulations were performed for progressively increasing values of \(C_{\mathcal{S}}^{\mathrm{res}}/C_{\mathcal{A}}^{\mathrm{res}}\). 

\begin{figure}
    \centering
    \subfloat[\label{fig:migration_flux_ratio}]{
        \includegraphics[width=\linewidth]
        {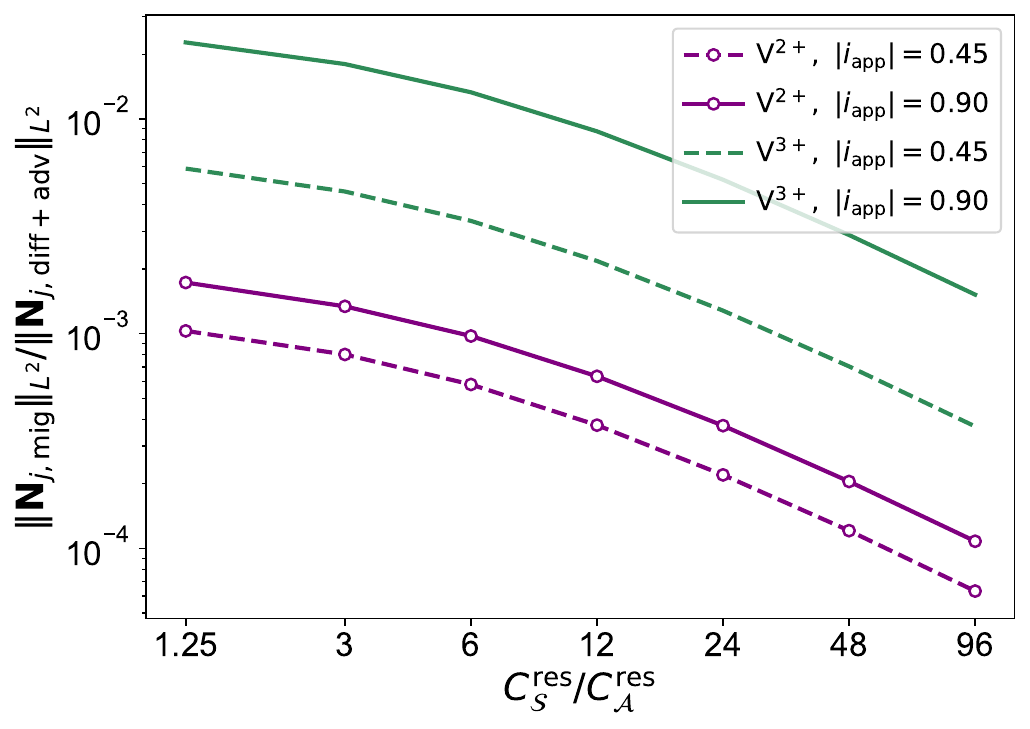}
    }
    \par\medskip   
    \subfloat[\label{fig:supporting_electrolyte_relative_error}]{
        \includegraphics[width=\linewidth]
        {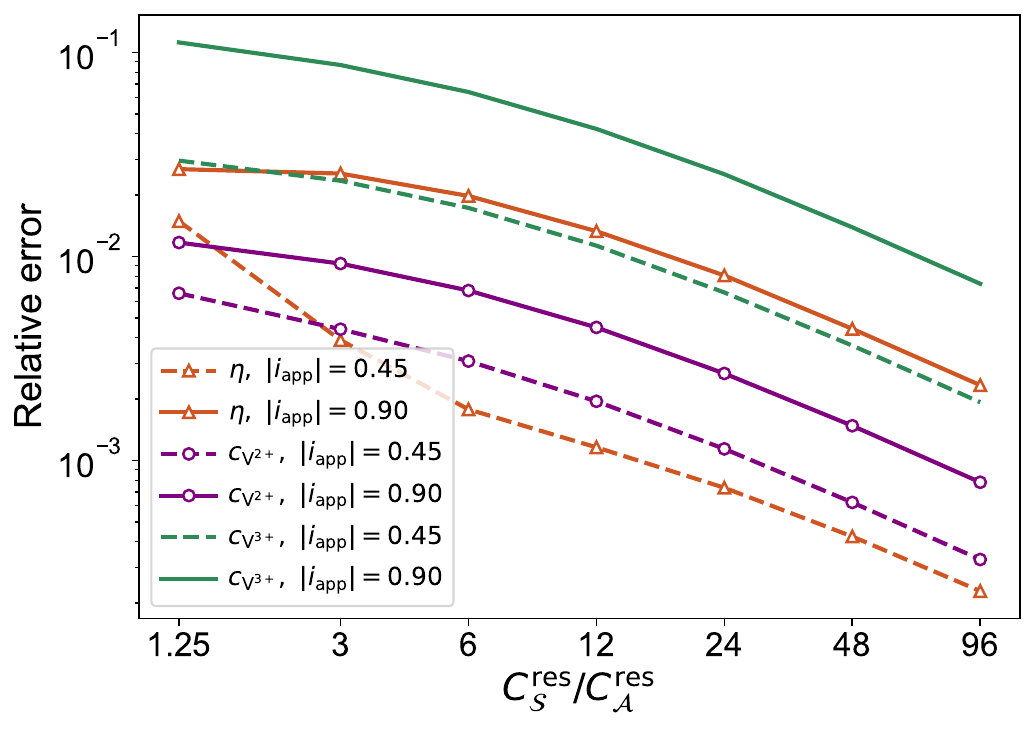}
    }

    \caption{
    Influence of the reservoir supporting-to-active-electrolyte concentration ratio, \(C_{\mathcal{S}}^{\mathrm{res}}/C_{\mathcal{A}}^{\mathrm{res}}\), on the approach to the ESE limit.
    (a) Ratio of the \(L^2\) norm of the electromigration flux to that of the combined diffusion and advection fluxes for \(\mathrm{V}^{2+}\) and \(\mathrm{V}^{3+}\), evaluated using the full model.
    (b) Relative \(L^2\) errors in \(\eta\), \(c_{\mathrm{V}^{2+}}\), and \(c_{\mathrm{V}^{3+}}\) between the reduced and full models, normalized by the corresponding full-model solutions.
    Results are shown for applied-current-density magnitudes \(\lvert i_{\mathrm{app}}\rvert=0.45\,\mathrm{A\,m^{-2}}\) and \(0.90\,\mathrm{A\,m^{-2}}\).}
    \label{fig:supporting_electrolyte_convergence}
\end{figure}

\Cref{fig:supporting_electrolyte_convergence}\mbox{(a)} shows the relative importance of electromigration in the redox-active-species fluxes predicted by the full model. 
For \(j\in\{\mathrm{V}^{2+},\mathrm{V}^{3+}\}\), we evaluate
\begin{equation}\label{eq:migration-contribution}
\frac{\|\bm N_{j,\mathrm{mig}}\|_{L^2(\Omega)}}
{\|\bm N_{j,\mathrm{diff}}+\bm N_{j,\mathrm{adv}}\|_{L^2(\Omega)}}.
\end{equation}
This quantity compares the magnitude of the electromigration flux with that of the combined diffusion and advection fluxes. 
We see that the ratio decreases for both species as \(C_{\mathcal S}^{\mathrm{res}}/C_{\mathcal A}^{\mathrm{res}}\) increases, confirming that electromigration becomes negligible in the ESE limit.
At a fixed concentration ratio, the ratio is larger at higher applied-current magnitudes; hence, larger potential gradients are required to sustain the imposed current and the resulting increase in electromigration.

\Cref{fig:supporting_electrolyte_convergence}\mbox{(b)} shows the relative \(L^2\) errors in the overpotential and redox-active species concentrations, computed using \cref{eq:relative-L2-error} over \(\mathcal D=\Omega\), with the reduced-model solution as \(q_a\) and the corresponding full-model solution as \(q_b\).
For small concentration ratios, the reduced and full formulations differ appreciably, particularly for \(c_{\mathrm{V}^{3+}}\).
At a fixed concentration ratio, the larger errors at higher applied-current magnitudes are consistent with stronger potential and concentration gradients, which increase the neglected electromigration fluxes and concentration-gradient contributions to the electrolyte current density, making the constant-conductivity approximation less accurate.
For all curves, as \(C_{\mathcal S}^{\mathrm{res}}/C_{\mathcal A}^{\mathrm{res}}\) increases, the relative errors decrease, demonstrating that the reduced formulation approaches the full model in the ESE limit.

Having confirmed the validity of the reduced model in the ESE limit, next, we use the reduced formulation for further verification studies: a 1D comparison with a high-accuracy ODE solution and PyBaMM, and a 3D comparison with RfbFoam.

\subsubsection{Verification and performance comparison in ESE limit}

\paragraph{One-dimensional comparison}\mbox{}

To further assess \software{} against an established open-source battery-modeling framework, we considered a 1D problem in the ESE limit and implemented it using the 1D version of \software{} and PyBaMM~\cite{Sulzer2021}.
PyBaMM is an open-source Python framework for battery simulation. 
Rather than using one of its predefined models, we constructed a custom model with \texttt{pybamm.BaseModel}, directly specifying the state variables, governing equations, and boundary conditions.
Specifically, we considered a steady-state problem without electrolyte flow, assuming that surface and bulk redox-active-species concentrations are equal. 
The redox-active-species fluxes, therefore, reduce to diffusion fluxes.
The resulting boundary-value problem is governed by the following equations:
\begin{subequations}\label{eq:ESE_model2}
\begin{align}
-&\partial_x
\left(
D_j^{\mathrm{eff}}\partial_x c_j
\right)
=
\begin{cases}
-\dfrac{a i_F}{F},
& j=\mathrm{V}^{2+},
\\[6pt]
+\dfrac{a i_F}{F},
& j=\mathrm{V}^{3+},
\end{cases}
\label{eq:ESE_model2_species}
\\[6pt]
&\partial_x
\left(
-\kappa_{\mathrm{ESE}}^{\mathrm{eff}} \partial_x\phi_\ell
\right)
=
a i_F,
\label{eq:ESE_model2_liquid}
\\[6pt]
&\partial_x
\left(
-\sigma^{\mathrm{eff}}\partial_x\phi_s
\right)
=
-a i_F,
\label{eq:ESE_model2_solid}
\\[6pt]
&i_F
=
i_0
\left[
\exp\!\left(
\frac{\alpha_a F}{RT}\eta
\right)
-
\exp\!\left(
-\frac{\alpha_c F}{RT}\eta
\right)
\right].
\label{eq:ESE_model2_BV}
\end{align}
\end{subequations}
We used the relevant parameter values listed in \cref{tab:representative_physical_params}.
For this comparison, we retained the zero-flux boundary condition for the redox-active species at the current collector while prescribing their concentrations at the membrane:
\begin{equation}
c_{\mathrm{V}^{2+}}
=
c_{\mathrm{V}^{3+}}
=
300~\mathrm{mol\,m^{-3}}
\qquad
\text{on } \Gamma_{\mathrm{mem}}.
\label{eq:bc-concentration-1d-right}
\end{equation}

\begin{figure}
  \centering
    \subfloat[\label{fig:1d_compare_overpotential}]{
        \includegraphics[width=\linewidth]
        {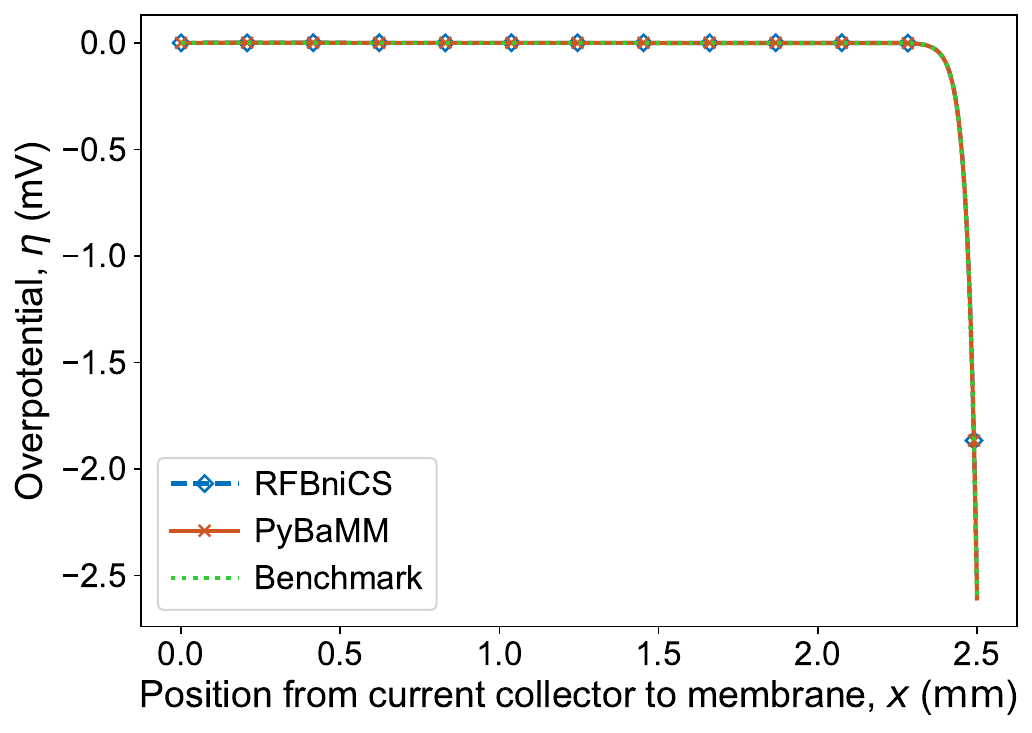}
    }
    \par\medskip
    
    \subfloat[\label{fig:1d_compare_concentration}]{
        \includegraphics[width=\linewidth]
        {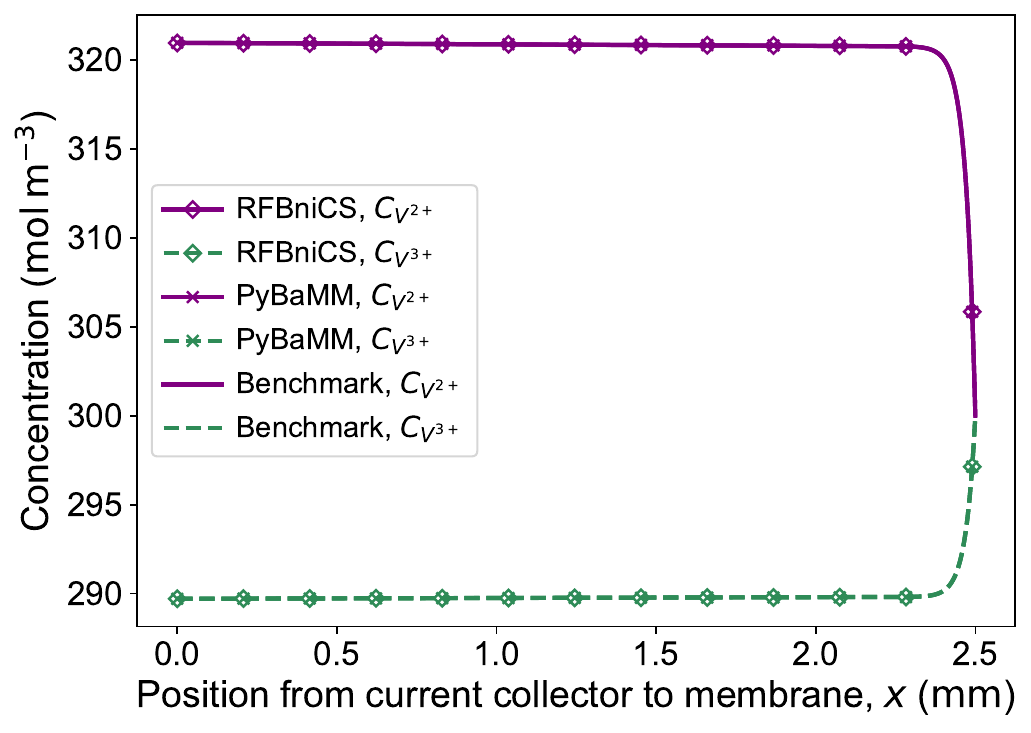}
    }

  \caption{Benchmark, PyBaMM, and \software{} solutions for the 1D problem in the excess-supporting-electrolyte limit: (a) overpotential profile and (b) redox-active-species concentration profiles.}
  
  \label{fig:excess_supporting_electrolyte_profiles}
\end{figure}

We solved the same boundary-value problem using PyBaMM, the 1D version of \software{}, and \texttt{scipy.integrate.solve\_bvp}. 
The latter method provided a high-resolution benchmark solution with which we assessed the accuracy of the PyBaMM and \software{} solutions.
\Cref{fig:excess_supporting_electrolyte_profiles} shows the benchmark, PyBaMM, and \software{} solutions for the overpotential profile in panel~(a) and the vanadium-ion concentration profiles in panel~(b).
We see that PyBaMM and \software{} closely reproduce the benchmark profiles. 
For the charging case considered here, the consumption of \(\mathrm{V}^{3+}\) and production of \(\mathrm{V}^{2+}\) cause their concentrations to decrease and increase, respectively, while their prescribed values remain fixed at the membrane.

\begin{figure}
  \centering
    \subfloat[\label{fig:1d_compare_l2_error}]{
        \includegraphics[width=\linewidth]
        {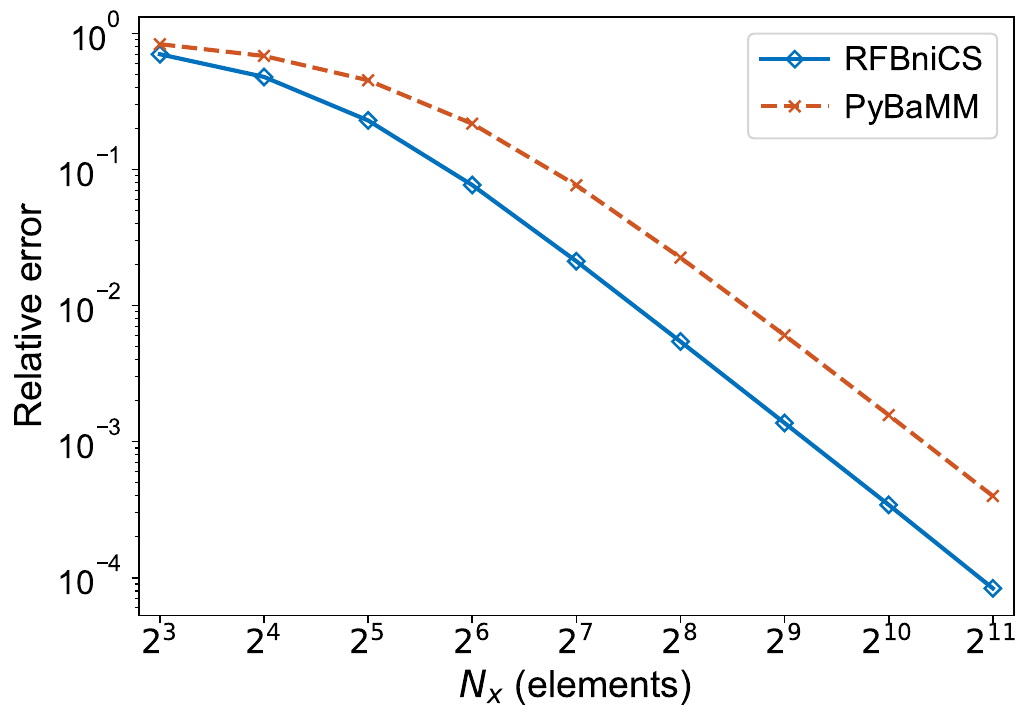}
    }
    \par\medskip
    
    \subfloat[\label{fig:1d_compare_runtime}]{
        \includegraphics[width=\linewidth]
        {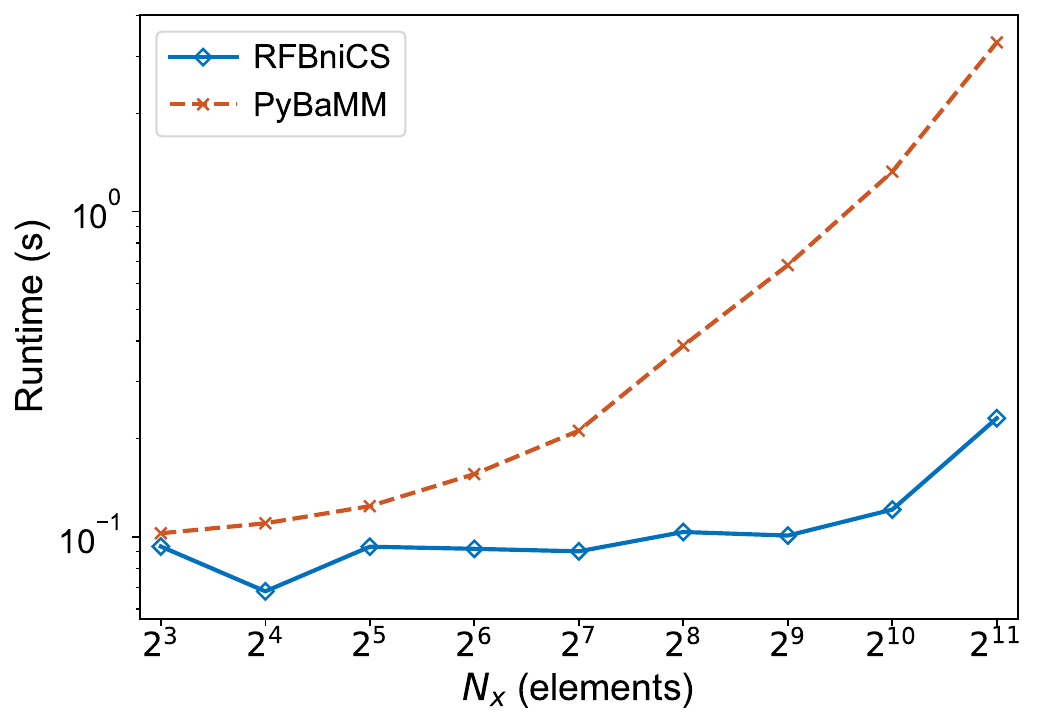}
    }

  \caption{Accuracy and runtime of the PyBaMM and 1D \software{} implementations for the problem in the excess-supporting-electrolyte limit as functions of spatial resolution: (a) relative \(L^2\) error in the overpotential with respect to the high-resolution benchmark and (b) wall-clock runtime.}

  \label{fig:pybamm_comparison}
\end{figure}

We quantified the agreement with the benchmark using the relative \(L^2\) error defined in~\cref{eq:relative-L2-error}, with \(\mathcal D=(0,W)\).
We then repeated the refinement study to examine how the error changes with the spatial resolution in the \(x\)-direction. 
\Cref{fig:pybamm_comparison}\mbox{(a)} shows the resulting overpotential errors, indicating that \software{} is more accurate than PyBaMM across all tested resolutions.
We observed similar trends for \(c_{\mathrm{V}^{2+}}\) and \(c_{\mathrm{V}^{3+}}\) (not shown). 
The runtime comparison in \cref{fig:pybamm_comparison}\mbox{(b)} also shows that \software{} is generally faster than PyBaMM, and that the difference in runtime grows with increasing resolution.
Thus, \software{} achieves higher accuracy at a lower computational cost for this comparison.

\paragraph{Three-dimensional comparison}\label{RFBfoam}\mbox{}

\begin{table}
\centering
\caption{Parameter values used for the comparison with RfbFoam. 
Except for the inlet velocity, the values correspond to the FTFF scenario provided in the RfbFoam repository.\footnote{URL: \url{https://github.com/mbarzegary/RfbFoam}} 
}
\label{tab:rfbfoam_comparison_parameters}
\scriptsize
\setlength{\tabcolsep}{4pt}
\renewcommand{\arraystretch}{1.15}
\begin{tabular}{@{}l l@{}}
\toprule
Quantity & Value (unit) \\
\midrule
\multicolumn{2}{@{}l@{}}{\textit{Geometry and mesh}} \\
Porous-electrode extent in \(x\), \(W\)
& \(0.42\ \mathrm{mm}\) \\
Porous-electrode extent in \(y\), \(L\)
& \(17.0\ \mathrm{mm}\) \\
Porous-electrode extent in \(z\), \(H\)
& \(15.0\ \mathrm{mm}\) \\
Channel extent in \(x\), \(W_{\mathrm{ch}}\)
& \(1.0\ \mathrm{mm}\) \\
Channel extent in \(y\), \(L_{\mathrm{ch}}\)
& \(1.0\ \mathrm{mm}\) \\
Rounded-channel end-cap radius, \(r_{\mathrm{ch}}\)
& \(0.5\ \mathrm{mm}\) \\
Channel centerline start in \(z\), \(z_{\mathrm{ch,start}}\)
& \(1.5\ \mathrm{mm}\) \\
Channel centerline end in \(z\), \(z_{\mathrm{ch,end}}\)
& \(13.5\ \mathrm{mm}\) \\
Total rounded-channel span in \(z\), \(H_{\mathrm{ch}}\)
& \(13.0\ \mathrm{mm}\) \\
Inlet-channel center in \(y\), \(y_{\mathrm{in}}\)
& \(1.0\ \mathrm{mm}\) \\
Outlet-channel center in \(y\), \(y_{\mathrm{out}}\)
& \(16.0\ \mathrm{mm}\) \\
Inlet/outlet patch area, \(A_{\mathrm{in}}=A_{\mathrm{out}}\)
& \(12.785\ \mathrm{mm^2}\) \\
\midrule
\multicolumn{2}{@{}l@{}}{\textit{Flow and porous-medium parameters}} \\
Imposed inlet velocity, \(v_{\mathrm{in}}\)
& \(1.0\times10^{-7}\ \mathrm{m\,s^{-1}}\) \\
Electrolyte density, \(\rho\)
& \(1015.0\ \mathrm{kg\,m^{-3}}\) \\
Dynamic viscosity, \(\mu\)
& \(1.14289\times10^{-3}\ \mathrm{Pa\,s}\) \\
Porosity, \(\varepsilon\)
& \(0.877\) \\
Permeability, \(K\)
& \((9.18\times10^{10})^{-1}\ \mathrm{m^2}\) \\
Tortuosity, \(\tau\)
& \(1.42\) \\
Pore-length scale, \(L_p\)
& \(100\times10^{-6}\ \mathrm{m}\) \\
\midrule
\multicolumn{2}{@{}l@{}}{\textit{Electrochemical and transport parameters}} \\
Solid-phase conductivity, \(\sigma\)
& \(275.78\ \mathrm{S\,m^{-1}}\) \\
Electrolyte conductivity, \(\kappa\)
& \(34.84\ \mathrm{S\,m^{-1}}\) \\
Specific interfacial area, \(a\)
& \(68010.0\ \mathrm{m^{-1}}\) \\
Charge-transfer coefficients, \(\alpha_a=\alpha_c\)
& \(0.5\) \\
Exchange current density, \(i_0\)
& \(165.0\ \mathrm{A\,m^{-2}}\) \\
Equilibrium potential, \(E_{\mathrm{eq}}\)
& \(0.771\ \mathrm{V}\) \\
Temperature, \(T\)
& \(293.5\ \mathrm{K}\) \\
Number of transferred electrons, \(n_e\)
& \(1.0\) \\
Diffusivity of reduced species, \(D_\mathcal{R}\)
& \(5.7\times10^{-10}\ \mathrm{m^2\,s^{-1}}\) \\
Diffusivity of oxidized species, \(D_\mathcal{O}\)
& \(4.8\times10^{-10}\ \mathrm{m^2\,s^{-1}}\) \\
Inlet concentration of reduced species, \(c_\mathcal{R}^{\mathrm{in}}\)
& \(250.0\ \mathrm{mol\,m^{-3}}\) \\
Inlet concentration of oxidized species, \(c_\mathcal{O}^{\mathrm{in}}\)
& \(250.0\ \mathrm{mol\,m^{-3}}\) \\
Reference concentration, \(c_{\mathrm{ref}}\)
& \(250.0\ \mathrm{mol\,m^{-3}}\) \\
\midrule
\multicolumn{2}{@{}l@{}}{\textit{Mass-transfer correlation}} \\
Mass-transfer prefactor, \(g_{km}\)
& \(7.75\times10^{-4}\ \mathrm{m\,s^{-1}}\) \\
Mass-transfer exponent, \(b_{km}\)
& \(0.9818\) \\
Reference velocity, \(v_{\mathrm{ref}}\)
& \(1.0\ \mathrm{m\,s^{-1}}\) \\
\midrule
\multicolumn{2}{@{}l@{}}{\textit{Boundary conditions and solver settings}} \\
Solid potential at current collector, \(\phi_s|_{\Gamma_{\mathrm{cc}}}\)
& \(0.0\ \mathrm{V}\) \\
Electrolyte potential at membrane, \(\phi_\ell|_{\Gamma_{\mathrm{mem}}}\)
& \(-1.0\ \mathrm{V}\) \\
\bottomrule
\end{tabular}
\end{table}

Next, we compared predictions of \software{} with those of RfbFoam, which solves RFB problems in the ESE approximation~\cite{Barzegari2026RfbFoam}.
The RfbFoam repository includes an example with a 3D FTFF geometry, the \(\mathrm{Fe}^{2+}/\mathrm{Fe}^{3+}\) redox couple, and electrochemical parameters and transport properties listed in~\cref{tab:rfbfoam_comparison_parameters}.
Notably, RfbFoam uses three constitutive relations that differ from the \software{} formulation discussed so far, and which we changed in this subsection only, to facilitate a direct comparison. 
First, RfbFoam evaluates the interfacial Faradaic current density using the following expression instead of \cref{eq:i_F_VRFB_expanded}:
\begin{equation}
\begin{aligned}
&i_F
=\\ 
&\frac{
i_0
\left[
\frac{c_\mathcal{R}}{c_{\mathrm{ref}}}
\exp\left(\frac{\alpha_a F}{RT}\eta\right)
-
\frac{c_\mathcal{O}}{c_{\mathrm{ref}}}
\exp\left(-\frac{\alpha_c F}{RT}\eta\right)
\right]
}{
1
+
\frac{i_0}{n_e F k_{\mathcal{R}} c_{\mathrm{ref}}}
\exp\left(\frac{\alpha_a F}{RT}\eta\right)
+
\frac{i_0}{n_e F k_{\mathcal{O}} c_{\mathrm{ref}}}
\exp\left(-\frac{\alpha_c F}{RT}\eta\right)
}.
\label{eq:rfbfoam_bv}
\end{aligned}
\end{equation}
This formulation eliminates the surface concentrations using a linear film mass-transfer approximation. 

Second, instead of the Bruggeman-type relation for the effective diffusivity~\eqref{eq:Bruggeman}, RfbFoam employs 
\begin{equation}
D_j^{\mathrm{eff}}
=
D_j \frac{\varepsilon}{\tau},
\label{eq:rfbfoam_deff}
\end{equation}
where \(\tau\) denotes the tortuosity factor (--).

Third, instead of the mass-transfer correlation in~\cref{eq:mass_transfer_km}, RfbFoam employs 
\begin{equation}
k_{\mathcal{R}} = k_{\mathcal{O}}
=
g_{km}
\left(
\frac{\lvert\bm{v}\rvert}{v_{\mathrm{ref}}}
\right)^{b_{km}},
\label{eq:rfbfoam_mass_transfer_km}
\end{equation}
where \(g_{km}\) is the mass-transfer prefactor \((\mathrm{m\,s^{-1}})\), \(b_{km}\) is the velocity exponent (--), and \(v_{\mathrm{ref}}\) is the reference velocity.

To ensure a consistent and direct comparison between \software{} and RfbFoam, we adopted the 3D FTFF geometry, potentiostatic boundary conditions, and the RfbFoam-specific constitutive relations in~\cref{eq:rfbfoam_bv,eq:rfbfoam_deff,eq:rfbfoam_mass_transfer_km}.
The parameter values were also taken from the corresponding repository case, except for the inlet velocity. 
For the present comparison, a lower inlet velocity was used in both implementations to keep the local mesh Péclet number below the recommended stability threshold.
This reduces the likelihood that mesh-induced numerical artifacts associated with advection-dominated transport contribute to differences between the \software{} and RfbFoam solutions.

\begin{figure}
  \centering
    \subfloat[\label{fig:rfbfoam_concentration_comparison}]{
        \includegraphics[width=\linewidth]
        {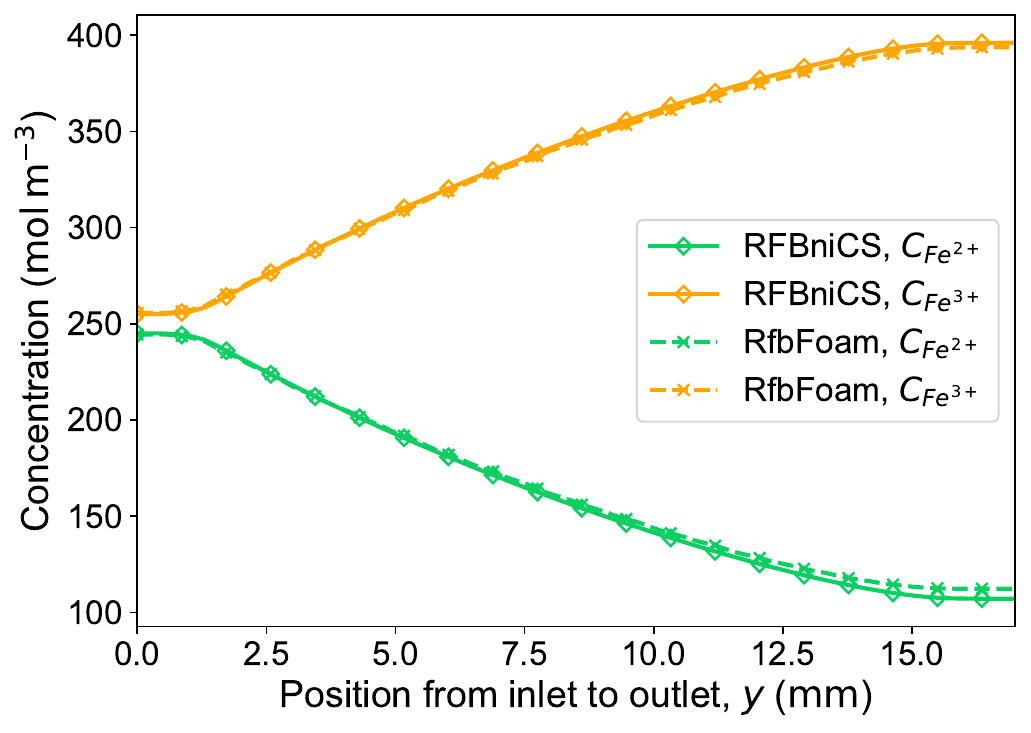}
    }
    \par\medskip
    
    \subfloat[\label{fig:rfbfoam_overpotential_comparison}]{
        \includegraphics[width=\linewidth]
        {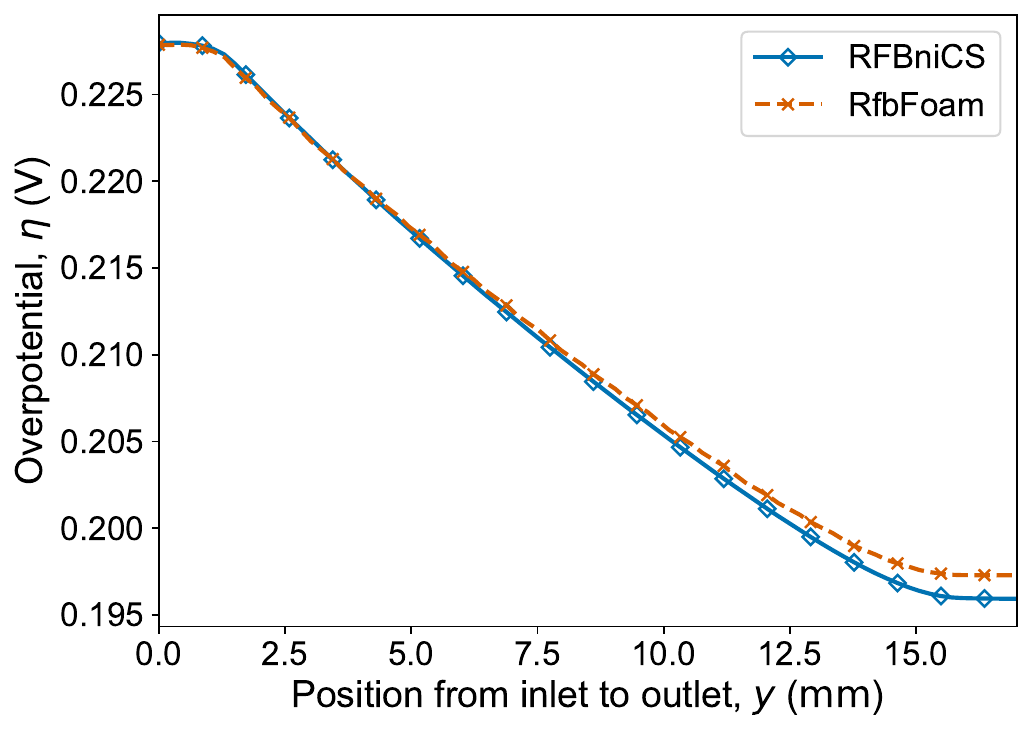}
    }
    \caption{Comparison of cross-sectionally averaged profiles obtained with \software{} and RfbFoam: (a) \(\mathrm{Fe}^{2+}\) and \(\mathrm{Fe}^{3+}\) concentrations and (b) overpotential.}
  \label{fig:rfbfoam_comparison}
\end{figure}

\software{} and RfbFoam simulations of the FTFF geometry yield 3D profiles of the overpotential and redox-active-species concentrations. 
To facilitate a comparison between the models, we averaged the 3D profiles over the transverse \(x\)- and \(z\)-directions within the porous-electrode region, yielding 1D profiles varying along the  \(y\)-axis, that is, the inlet-to-outlet direction.
The resulting cross-sectionally averaged concentration and overpotential profiles are shown in \cref{fig:rfbfoam_comparison}.
Their close agreement demonstrates that \software{} reproduces the RfbFoam FTFF solution for the matched model configuration, thereby providing another verification for \software{}.
The deviation between the full 3D fields was quantified using the relative \(L^2\) error as defined in \cref{eq:relative-L2-error}, with \(\mathcal D=\Omega\), where the \software{} field was used as \(q_a\), and the corresponding RfbFoam field was used as the reference quantity \(q_b\).
The resulting relative \(L^2\) errors were $1.7\%$ for \(c_{\mathrm{Fe}^{2+}}\), $0.54\%$ for \(c_{\mathrm{Fe}^{3+}}\), and $0.39\%$  for the overpotential \(\eta\).

\subsection{Representative results from \software}

Following the verification and comparison studies presented above, we use \software{} to simulate transient operation of the negative half-cell of a VRFB under conditions not accessible to the other open-source RFB frameworks considered in this study.
The simulation accounts for the spatiotemporal variation of the supporting-electrolyte species and is performed in the 2D geometry using the baseline parameters listed in~\cref{tab:representative_physical_params}. 
The operating protocol consists of galvanostatic charging for \(160112\ \mathrm{s}\), a rest period of \(67105\ \mathrm{s}\), and galvanostatic discharging for \(139112\ \mathrm{s}\).
During the intermediate rest period, the current is set to zero while electrolyte flow is retained.
The negative half-cell voltage is evaluated as the difference between the boundary-averaged solid-phase potential at the current collector and the boundary-averaged electrolyte-phase potential at the membrane.

\begin{figure}[t]
  \centering

  \subfloat[\label{fig:representative-cycle-voltage}]{
    \includegraphics[width=\linewidth]
    {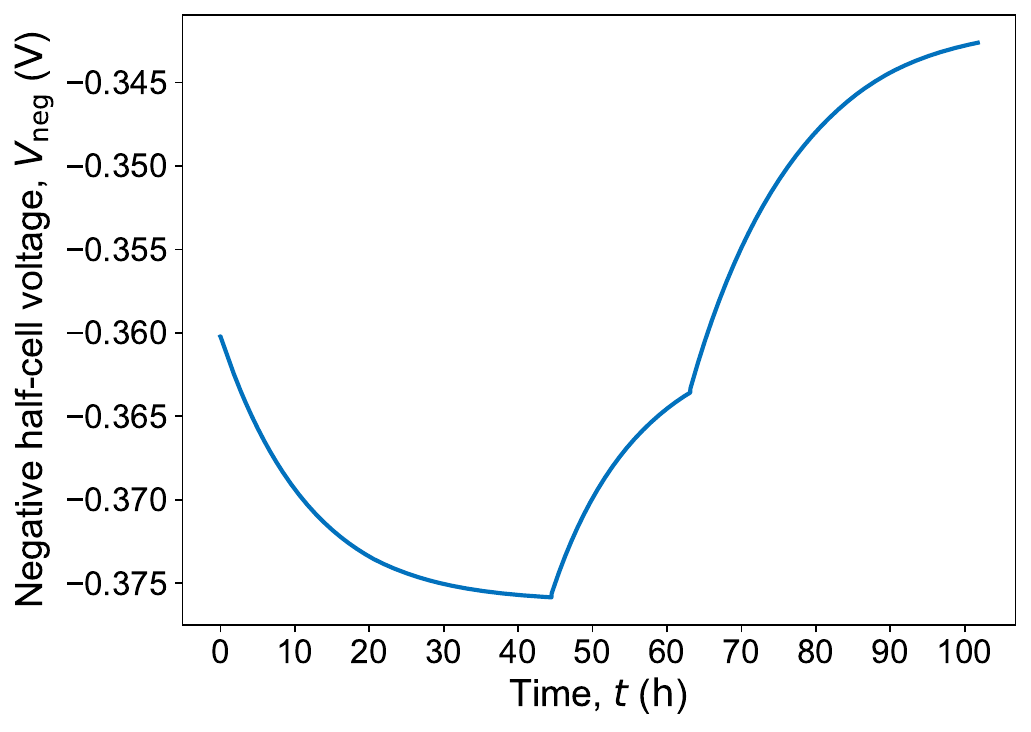}
  }

  \par\medskip

  \subfloat[\label{fig:representative-cycle-concentrations}]{
    \includegraphics[width=\linewidth]
    {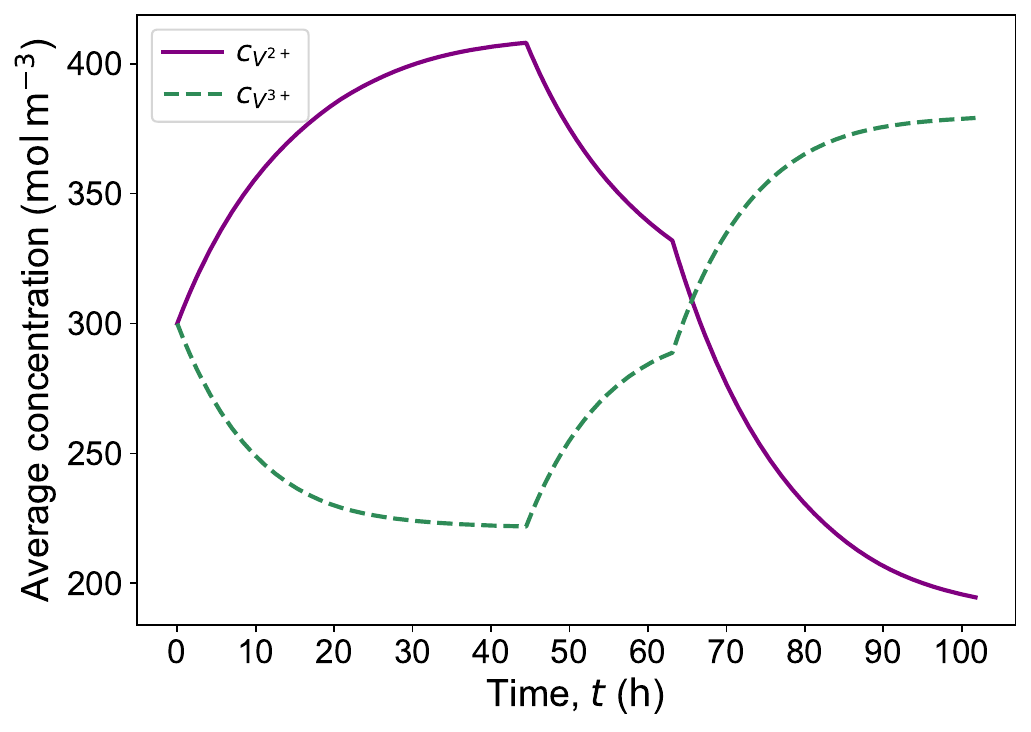}
  }

  \caption{Transient response during one galvanostatic
  charge--rest--discharge cycle:
  (a) negative half-cell voltage and
  (b) spatially averaged concentrations of
  \(\mathrm{V}^{2+}\) and \(\mathrm{V}^{3+}\).}
  \label{fig:representative-cycle}
\end{figure}

\Cref{fig:representative-cycle} shows the negative half-cell voltage in panel~(a) and the corresponding spatially averaged concentrations of
\(\mathrm{V}^{2+}\) and \(\mathrm{V}^{3+}\) in panel~(b).
During charging, the magnitude of the voltage increases as \(\mathrm{V}^{3+}\) is converted to \(\mathrm{V}^{2+}\). 
At the beginning of the rest period, when the applied current is set to zero, \(V_{\mathrm{neg}}\) reverses direction and increases toward less
negative values.
During the rest period, the concentrations change, no longer because of Faradaic reactions, but mainly due to the maintained electrolyte flow.
During discharge, the voltage magnitude decreases gradually, and \(\mathrm{V}^{2+}\) is converted to \(\mathrm{V}^{3+}\).

\begin{figure}
  \centering
    \subfloat[ \label{fig:representative-electrolyte-phase-potential}]{
        \includegraphics[width=\linewidth]
        {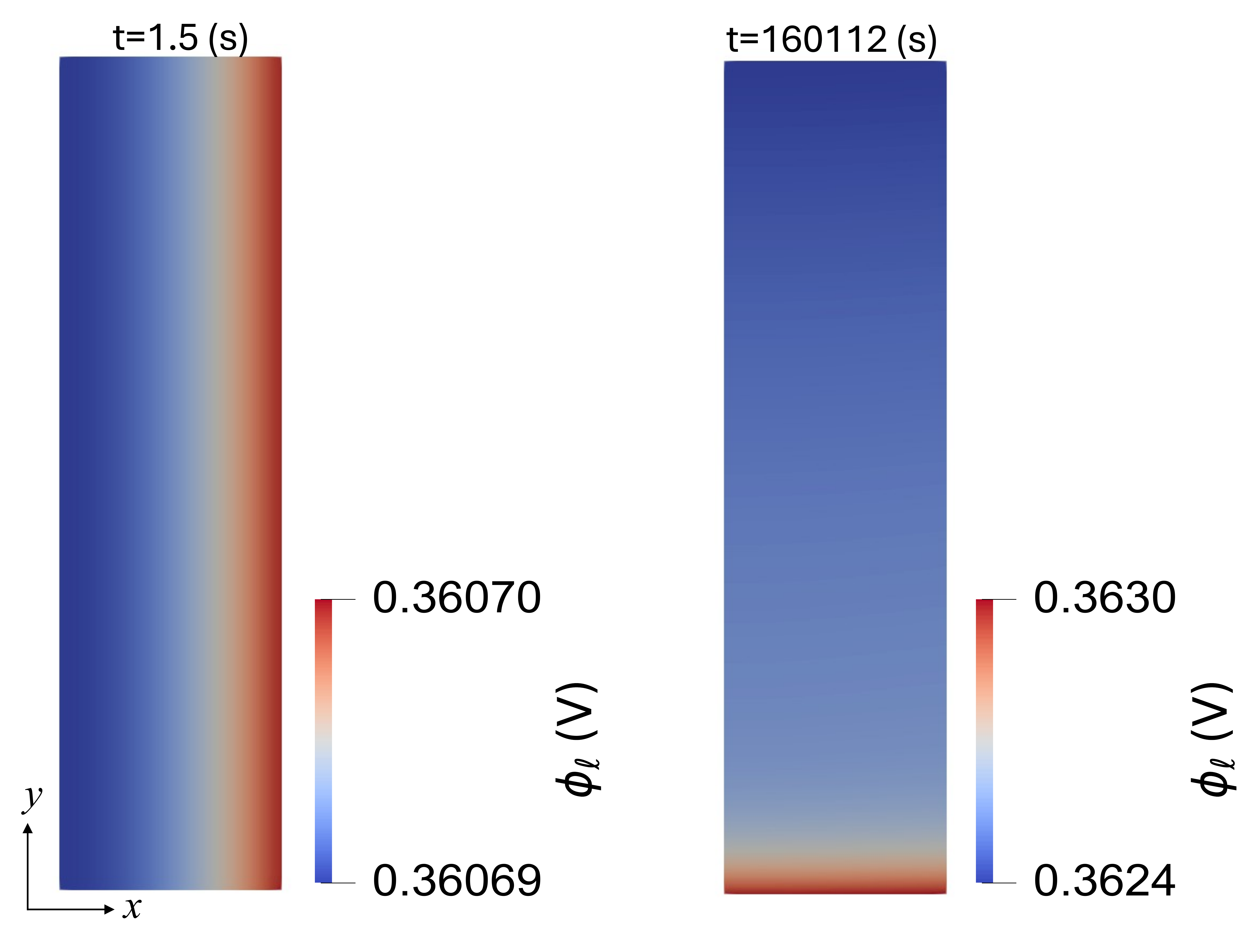}
    }
    \par\medskip
    
    \subfloat[\label{fig:representative-solid-phase-potential}]{
        \includegraphics[width=\linewidth]
        {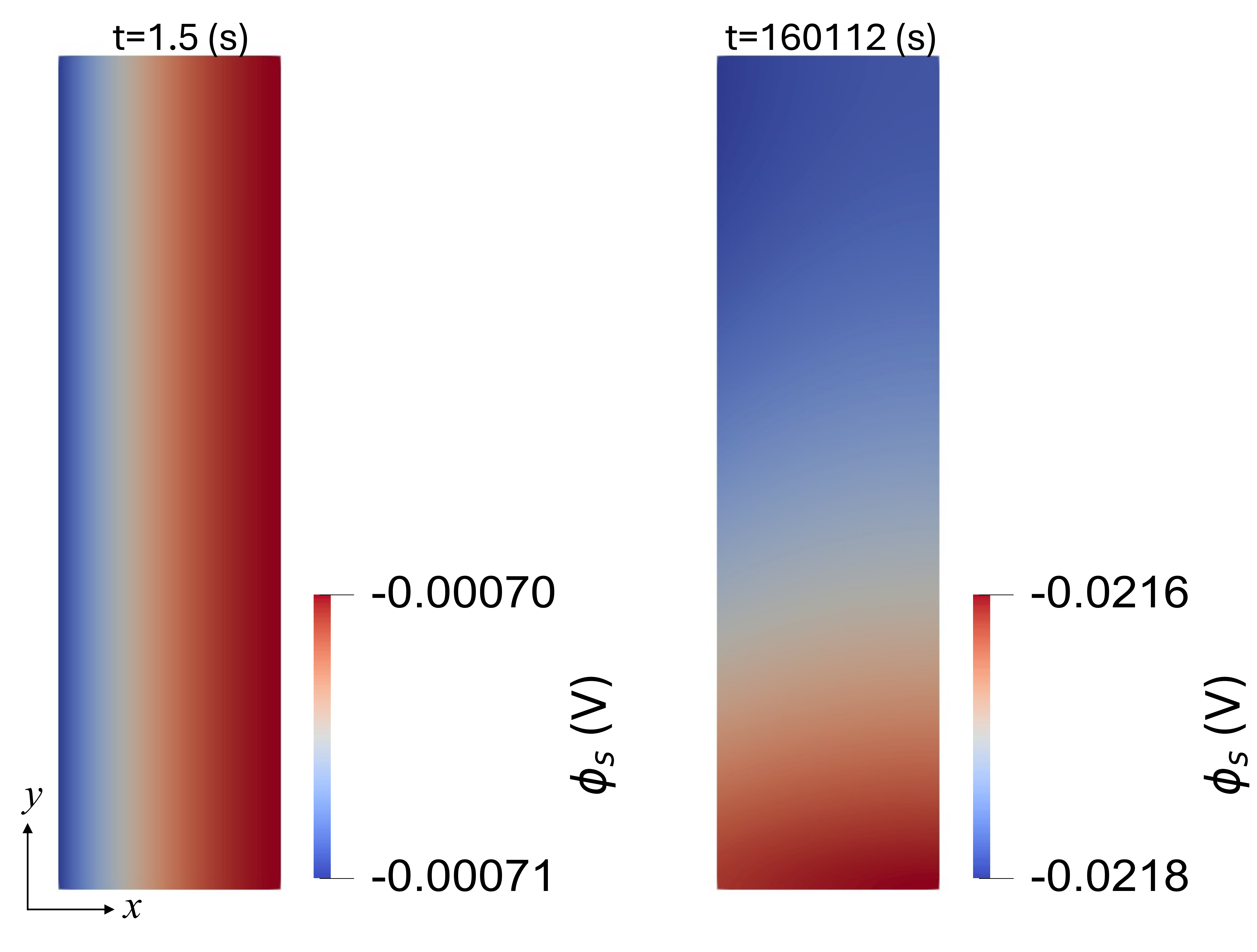}
    }
        
    \par\medskip
    
    \subfloat[\label{fig:representative-faradaic-current}]{
        \includegraphics[width=\linewidth]
        {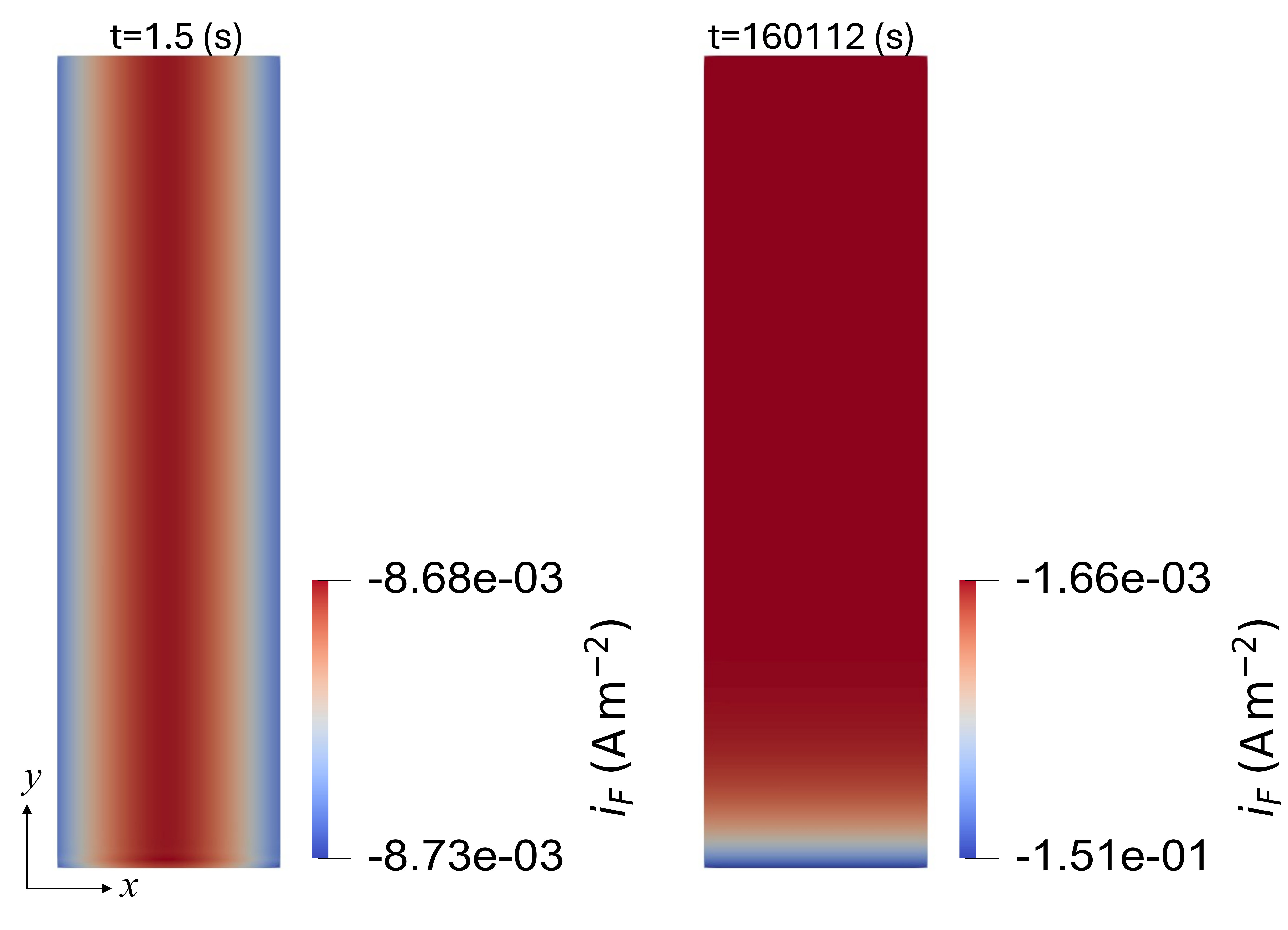}
    }
    \caption{Spatial distributions of (a) the electrolyte-phase potential, \(\phi_\ell\), (b) the solid-phase potential, \(\phi_s\), and (c) the interfacial Faradaic current density, \(i_F\), during galvanostatic charge at \(t=1.5\ \mathrm{s}\) (left) and \(t=160112\ \mathrm{s}\) (right).}
  \label{fig:representative-transient-fields}
\end{figure}

\Cref{fig:representative-transient-fields} compares the spatial distributions of the electrolyte-phase potential, solid-phase potential, and interfacial Faradaic current density at the beginning (\(t=1.5\ \mathrm{s}\)) and near the end (\(t=160112\ \mathrm{s}\)) of galvanostatic charging. The left and right columns correspond to the earlier and later times, respectively.
At \(t=1.5\ \mathrm{s}\), the species concentrations remain close to their initially uniform values. 
Consequently, concentration-dependent quantities such as the equilibrium potential, exchange current density, and electrolyte conductivity vary only weakly along the flow direction. 
The electrolyte- and solid-phase potentials in \cref{fig:representative-transient-fields}\mbox{(a)} and (b) therefore vary mainly in the through-plane direction, while their variation along the flow direction remains small.
The difference between these potentials, corrected by the local equilibrium potential, determines the activation overpotential and hence the local Faradaic current density through the Butler--Volmer relation. 
Accordingly, the \(i_F\) distribution in \cref{fig:representative-transient-fields}\mbox{(c)} is also nearly invariant along the flow direction at this early time, although some variation remains across the electrode thickness.

Near the end of the charging phase (at \(t=160112\ \mathrm{s}\)), reaction and electrolyte transport have produced appreciable concentration gradients along the flow direction. 
The resulting spatial variations in concentration-dependent properties produce corresponding variations in \(\phi_\ell\), \(\phi_s\), and \(i_F\), with the Faradaic current density becoming increasingly localized near the inlet.
Although the local distribution of the Faradaic current density changes, its integral over the porous-electrode region remains constrained by the applied current.
This spatial variation indicates non-uniform utilization of the porous electrode.
These results demonstrate the importance of resolving temporal changes in the electrochemical fields and highlight the ability of \software{} to perform transient simulations.

The simulations also illustrate that \software{} gives access to spatially resolved distributions of species concentrations, solid- and electrolyte-phase potentials, and the interfacial Faradaic current density \(i_F\)---quantities that are generally difficult to measure directly in experiments.

In addition to the quantities shown here, \software{} can be used to define and visualize other primary and derived fields, including the state of charge, equilibrium potential, supporting-electrolyte concentrations, surface concentrations of the redox-active species, ratios between surface and bulk concentrations, electrolyte pressure and velocity, overpotential, and other user-defined expressions. 
Such information can be used to identify concentration polarization, non-uniform current distribution, high-overpotential regions, and inefficient electrolyte utilization.
The framework is therefore useful for analyzing the physical mechanisms that control half-cell performance.

\section{Conclusion}\label{sec:conclusions}

This work presented \software, an open-source finite-element framework to simulate RFB half-cells using FEniCSx.
We implemented an established macro-homogeneous porous-electrode model for the coupled processes governing half-cell operation, including electrolyte flow, multicomponent ionic transport, solid- and electrolyte-phase charge conduction, and interfacial Faradaic charge transfer.
\software{} offers a verified and extensible basis for adding further physics, chemistries, constitutive relations, and geometries.

We verified the accuracy and capabilities of \software{} by comparing its predictions to those of other packages, for the limiting parameter regimes for which these packages were devised.
In the uniform-concentration charge-conservation limit, we compared \software{} against an independent, publicly available implementation and a high-resolution one-dimensional benchmark, assessing the implementation of coupled charge balances and Butler--Volmer kinetics.
Then, in the excess-supporting-electrolyte limit, we assessed one-dimensional \software{} and PyBaMM implementations against a high-resolution benchmark and compared \software{} with RfbFoam for an equivalent three-dimensional FTFF case, thereby assessing the implementation of redox-active-species transport and its coupling to charge conservation, Faradaic kinetics, and electrolyte flow.
Together, these studies support the numerical consistency and accuracy of the corresponding components of the full \software{} implementation.

Beyond the limiting-regime comparisons, we demonstrated the broader capabilities of \software{} through a transient galvanostatic charge--rest--discharge simulation of a VRFB negative half-cell with a moderately concentrated supporting electrolyte. 
This case resolved the spatiotemporal evolution of both redox-active and supporting-electrolyte species beyond the excess-supporting-electrolyte approximation.

The \software{} results should be interpreted in light of the model assumptions. 
Overall, these assumptions enable efficient simulation of the main half-cell transport and electrochemical processes, while leaving more detailed descriptions for future extensions.

This research could be extended to account for thermal effects, multi-electron reactions, side reactions, degradation mechanisms, concentration-dependent material properties, alternative redox chemistries, and other flow-field configurations. 
In addition, the present half-cell formulation could be developed into a coupled full-cell model that explicitly accounts for membrane transport, including the crossover of redox-active species and the movement of water between the two half-cells. 
The framework could further be coupled to a storage-tank model with finite electrolyte volume, allowing the inlet concentrations to vary with the evolving tank composition.
Future work could also consider stabilized finite-element schemes for transport dominated by advection and electromigration. 
One possible approach is a discontinuous Galerkin formulation with an upwind treatment of the combined species drift and a weighted interior-penalty treatment of diffusion~\cite{ERN2009235}.

\section*{Acknowledgment}
The authors acknowledge funding from the EU's Horizon Europe research and innovation program under grant agreement \textit{no.}~101137725 (BatCAT).

\section*{Code availability}
The source code for RFBniCS and the scripts used for the comparison studies presented in this work are available at
\url{https://github.com/Ah-Aghabarari/RFBniCS}.

\section*{Author Declarations}
The authors have no conflict of interest to disclose.
During software development, ChatGPT (OpenAI) was used to assist with code refactoring and to improve code readability. 
The authors reviewed and tested all AI-assisted code modifications.

\onecolumngrid
\appendix

\numberwithin{equation}{section}
\renewcommand{\theequation}{\thesection.\arabic{equation}}

\numberwithin{figure}{section}
\renewcommand{\thefigure}{\thesection.\arabic{figure}}

\numberwithin{table}{section}
\renewcommand{\thetable}{\thesection.\arabic{table}}

\section{Derivation of the surface-concentration relations \eqref{eq:surface_concentration}}
\label{app:surface-concentrations}

Substituting the exchange-current-density expression~\eqref{eq:exchange_current_density} into the Butler--Volmer equation~\eqref{eq:butler_volmer} and dividing by \(F\), the local molar Faradaic reaction rate can be written as
\begin{align}
\frac{i_F}{F}
&=
k^0 (c_\mathcal{R})^{\alpha_c}(c_\mathcal{O})^{\alpha_a}
\left[
\frac{c_\mathcal{R}^{\mathrm{surf}}}{c_\mathcal{R}}
\exp\left(\frac{\alpha_a F\eta}{RT}\right)
-
\frac{c_\mathcal{O}^{\mathrm{surf}}}{c_\mathcal{O}}
\exp\left(-\frac{\alpha_c F\eta}{RT}\right)
\right]
\nonumber\\
&=
\underbrace{k^0 (c_\mathcal{R})^{\alpha_c-1}(c_\mathcal{O})^{\alpha_a}
\exp\left(\frac{\alpha_a F\eta}{RT}\right)}_{\equiv A}c_\mathcal{R}^{\mathrm{surf}}
-
\underbrace{k^0 (c_\mathcal{R})^{\alpha_c}(c_\mathcal{O})^{\alpha_a-1}
\exp\left(-\frac{\alpha_c F\eta}{RT}\right)}_{\equiv B}c_\mathcal{O}^{\mathrm{surf}}.
\label{eq:if_appendix}
\end{align}
Substituting \cref{eq:if_appendix} into the
interfacial mass-transfer balance relation~\eqref{eq:interfacial_mass_transfer_balance} gives
\begin{subequations}
\begin{align}
k_{\mathcal{R}}(c_\mathcal{R} - c_\mathcal{R}^{\mathrm{surf}})
&=
A c_\mathcal{R}^{\mathrm{surf}}
-
B c_\mathcal{O}^{\mathrm{surf}},
\\
k_{\mathcal{O}}(c_\mathcal{O} - c_\mathcal{O}^{\mathrm{surf}})
&=
-A c_\mathcal{R}^{\mathrm{surf}}
+
B c_\mathcal{O}^{\mathrm{surf}} .
\end{align}
\end{subequations}
Rearranging gives the linear system
\begin{subequations}
\begin{align}
(k_{\mathcal{R}}+A)c_\mathcal{R}^{\mathrm{surf}}
-
B c_\mathcal{O}^{\mathrm{surf}}
&=
k_{\mathcal{R}}c_\mathcal{R},
\\
-A c_\mathcal{R}^{\mathrm{surf}}
+
(k_{\mathcal{O}}+B)c_\mathcal{O}^{\mathrm{surf}}
&=
k_{\mathcal{O}}c_\mathcal{O} .
\end{align}
\end{subequations}
Solving this system gives the surface-concentration expressions~\eqref{eq:surface_concentration}.

\section{Derivation of the weak form for the electrolyte-flow equations \eqref{eq:app-flow-final-weak-form_txt}}
\label{app:bulk-flow-weak-form}

In the porous-electrode region \(\Omega_E\), \cref{eq:momentum_nsdf,eq:continuity_NSDF} give the steady Navier--Stokes momentum balance with Darcy--Forchheimer resistance and the incompressibility constraint, respectively.

Let \(\bm w\) denote a velocity test function. Moving all terms in \cref{eq:momentum_nsdf} to the left-hand side and multiplying by \(\bm w\) and integrating over \(\Omega_E\) gives
\begin{align}
\int_{\Omega_E}
\frac{\rho}{\varepsilon^2}
\left(\bm v\cdot\nabla\right)\bm v\cdot\bm w\,dx
+
\int_{\Omega_E}
\nabla p\cdot\bm w\,dx
-
\int_{\Omega_E}
\frac{\mu}{\varepsilon}
\Delta\bm v\cdot\bm w\,dx
+
\int_{\Omega_E}
\frac{\mu}{K}\bm v\cdot\bm w\,dx
+
\int_{\Omega_E}
\rho\beta|\bm v|\bm v\cdot\bm w\,dx =0.
\label{eq:app-flow-tested-electrode}
\end{align}
Integrating the pressure-gradient and viscous terms by parts gives the following volume contribution from the porous-electrode region:
\begin{align}
&\int_{\Omega_E}
\frac{\rho}{\varepsilon^2}
\left(\bm v\cdot\nabla\right)\bm v\cdot\bm w\,dx
-
\int_{\Omega_E}
p\,\nabla\cdot\bm w\,dx
+
\int_{\Omega_E}
\frac{\mu}{\varepsilon}
\nabla\bm v:\nabla\bm w\,dx
+
\int_{\Omega_E}
\frac{\mu}{K}\bm v\cdot\bm w\,dx
+
\int_{\Omega_E}
\rho\beta|\bm v|\bm v\cdot\bm w\,dx .
\label{eq:app-flow-weak-electrode}
\end{align}

In the flow-channel region \(\Omega_F\), the porosity is set to \(\varepsilon=1\), and the porous-medium resistance terms are omitted by taking \(K^{-1}=0\) and \(\beta=0\). 
Therefore, the corresponding strong form in \(\Omega_F\) is
\begin{equation}
\rho
\left(\bm v\cdot\nabla\right)\bm v
+
\nabla p
-
\mu\Delta \bm v
=
\bm 0 .
\label{eq:app-flow-residual-channel}
\end{equation}
Repeating the same integration-by-parts procedure in \(\Omega_F\) gives the volume contribution
\begin{equation}
\int_{\Omega_F}
\rho
\left(\bm v\cdot\nabla\right)\bm v\cdot\bm w\,dx
+
\int_{\Omega_F}
\mu
\nabla\bm v:\nabla\bm w\,dx
-
\int_{\Omega_F}
p\,\nabla\cdot\bm w\,dx .
\label{eq:app-flow-weak-channel}
\end{equation}
The boundary terms vanish on boundaries where the velocity test function vanishes. The boundary contributions on the internal interface between \(\Omega_E\) and \(\Omega_F\) cancel when the velocity and normal traction are continuous across the interface. 
On the outlet boundary \(\Gamma_{\mathrm{out}}\subset\partial\Omega_F\), the velocity is not prescribed. 
The natural traction condition~\eqref{eq:bc-flow-outlet-traction} therefore gives the remaining outlet contribution
\begin{equation}
\int_{\Gamma_{\mathrm{out}}}
p_{\mathrm{out}}\,\bm n\cdot\bm w\,ds .
\label{eq:app-outlet-pressure-term}
\end{equation}

Multiplying the incompressibility equation~\eqref{eq:continuity_NSDF} by a pressure test function \(q\) and integrating over the full flow domain gives
\begin{equation}
\int_{\Omega_E\cup\Omega_F}
q\,\nabla\cdot\bm v\,dx
=
0 .
\label{eq:app-flow-continuity-weak}
\end{equation}

Introducing the finite-element approximations $\bm v \approx \bm v_h$, $p \approx p_h$, and the corresponding test functions \(\bm w_h\) and \(q_h\), the mixed weak form is obtained by combining the momentum residuals in \(\Omega_E\) and \(\Omega_F\) with the weak incompressibility constraint.
Thus, find \((\bm v_h,p_h)\in U_h\times Q_h\) such that, for all
\((\bm w_h,q_h)\in U_h\times Q_h\),
\begin{align}
&\int_{\Omega_E}
\frac{\rho}{\varepsilon^2}
\left(\bm v_h\cdot\nabla\right)\bm v_h\cdot\bm w_h\,dx
+
\int_{\Omega_F}
\rho
\left(\bm v_h\cdot\nabla\right)\bm v_h\cdot\bm w_h\,dx
-
\int_{\Omega_E\cup\Omega_F}
p_h\nabla\cdot\bm w_h\,dx
+
\int_{\Omega_E}
\frac{\mu}{\varepsilon}
\nabla\bm v_h:\nabla\bm w_h\,dx
\notag\\
&
+
\int_{\Omega_F}
\mu
\nabla\bm v_h:\nabla\bm w_h\,dx
+
\int_{\Omega_E}
\frac{\mu}{K}\bm v_h\cdot\bm w_h\,dx
+
\int_{\Omega_E}
\rho\beta |\bm v_h|\bm v_h\cdot\bm w_h\,dx
-
\int_{\Omega_E\cup\Omega_F}
q_h\nabla\cdot\bm v_h\,dx
+
\int_{\Gamma_{\mathrm{out}}}
p_{\mathrm{out}}\,\bm n\cdot\bm w_h\,ds
=0 .
\label{eq:app-flow-final-weak-form}
\end{align}

\section{Nondimensionalization of the governing equations}
\label{app:nondimensionalization}

\subsection{Nondimensionalization of the species mass balance}
We introduce the dimensionless variables
\begin{equation}
\begin{aligned}
\bm{x}^* &= \frac{\bm{x}}{L_{\mathrm{ref}}},
\quad
t^* = \frac{t}{t_{\mathrm{ref}}},
\quad
c_j^* = \frac{c_j}{c_{\mathrm{ref}}},
\quad
c_j^{\mathrm{surf},*} = \frac{c_j^{\mathrm{surf}}}{c_{\mathrm{ref}}},
\quad
D_j^{\mathrm{eff},*} = \frac{D_j^{\mathrm{eff}}}{D_{\mathrm{ref}}},
\\
\bm{v}^* &= \frac{\bm{v}}{U_{\mathrm{ref}}},
\quad
\phi_s^* = \frac{\phi_s}{\phi_{\mathrm{ref}}},
\quad
\phi_\ell^* = \frac{\phi_\ell}{\phi_{\mathrm{ref}}},
\quad
E_{\mathrm{eq}}^* = \frac{E_{\mathrm{eq}}}{\phi_{\mathrm{ref}}},
\quad
t_{\mathrm{ref}} = \frac{L_{\mathrm{ref}}}{U_{\mathrm{ref}}}.
\end{aligned}
\label{eq:dimensionless_variables}
\end{equation}

The Faradaic current density is scaled as
\begin{equation}
i_F=i_{\mathrm{ref}}i_F^*,
\qquad
i_{\mathrm{ref}}
=
F k^0 c_{\mathrm{ref}}^{\alpha_a+\alpha_c},
\end{equation}
where
\begin{equation}
i_F^*
=
(c_{\mathcal{R}}^*)^{\alpha_c}
(c_{\mathcal{O}}^*)^{\alpha_a}
\left[
\frac{c_{\mathcal{R}}^{\mathrm{surf},*}}{c_{\mathcal{R}}^*}
e^{\alpha_a \mathcal{M}_{\phi}\eta^*}
-
\frac{c_{\mathcal{O}}^{\mathrm{surf},*}}{c_{\mathcal{O}}^*}
e^{-\alpha_c \mathcal{M}_{\phi}\eta^*}
\right],
\qquad
\eta^*=\phi_s^*-\phi_\ell^*-E_{\mathrm{eq}}^* .
\end{equation}

Substitution of the dimensionless variables into the species mass balance~\eqref{eq:mass_balance} gives
\begin{equation}
\begin{aligned}
\frac{\partial(\varepsilon c_j^*)}{\partial t^*}
+
\nabla^*\cdot
\left[
-\frac{D_j^{\mathrm{eff},*}}{\mathrm{Pe}}\nabla^*c_j^*
-
\frac{z_j\mathcal{M}_{\phi}}{\mathrm{Pe}}
c_j^*D_j^{\mathrm{eff},*}\nabla^*\phi_\ell^*
+
\bm{v}^*c_j^*
\right]
&=
\mathrm{Da}_{c}
\begin{cases}
-i_F^*, & j=\mathcal{R},\\[6pt]
+i_F^*, & j=\mathcal{O},\\[6pt]
0, & j\in\mathcal{J}\setminus\{\mathcal{R},\mathcal{O},d\}.
\end{cases}
\end{aligned}
\label{eq:dimensionless_species_balance}
\end{equation}
The dimensionless numbers appearing in this equation are the Péclet number, the dimensionless potential scale, and the convective Damköhler number:
\begin{equation}
\mathrm{Pe}
=
\frac{U_{\mathrm{ref}}L_{\mathrm{ref}}}{D_{\mathrm{ref}}},
\qquad
\mathcal{M}_{\phi}
=
\frac{F\phi_{\mathrm{ref}}}{RT},
\qquad
\mathrm{Da}_{c}
=
\frac{a i_{\mathrm{ref}}L_{\mathrm{ref}}}{F c_{\mathrm{ref}}U_{\mathrm{ref}}}.
\end{equation}

\subsection{Nondimensionalization of the electrolyte charge balance}

Using the dimensionless variables defined above, the dimensionless electrolyte charge-balance equation becomes
\begin{equation}
\nabla^*\cdot
\left[
-\mathcal{M}_{\phi}
\left(
\sum_{j\in\mathcal{J}} z_j^2D_j^{\mathrm{eff},*}c_j^*
\right)
\nabla^*\phi_\ell^*
-
\sum_{j\in\mathcal{J}} z_jD_j^{\mathrm{eff},*}\nabla^*c_j^*
\right]
=
\mathrm{Pe}\,\mathrm{Da}_{c}\,i_F^* .
\label{eq:dimensionless_electrolyte_charge_balance}
\end{equation}

\subsection{Nondimensionalization of the solid-phase charge balance}

Using the dimensionless variables and Faradaic-current scaling defined above,
we introduce
\begin{equation}
\sigma^{\mathrm{eff},*}
=
\frac{\sigma^{\mathrm{eff}}}{\sigma_{\mathrm{ref}}}.
\end{equation}
The dimensionless solid-phase charge-balance equation is then
\begin{equation}
\nabla^*\cdot
\left(
-\sigma^{\mathrm{eff},*}\nabla^*\phi_s^*
\right)
=
-\delta_s i_F^*,
\label{eq:dimensionless_solid_charge_balance}
\end{equation}
with
\begin{equation}
\delta_s
=
\frac{a i_{\mathrm{ref}}L_{\mathrm{ref}}^2}
{\sigma_{\mathrm{ref}}\phi_{\mathrm{ref}}}.
\end{equation}
The parameter \(\delta_s\) compares interfacial Faradaic source strength with solid-phase ohmic conduction.

\subsection{Nondimensionalization of the electrolyte-flow equation}

The pressure is nondimensionalized as
\begin{equation}
p^*
=
\frac{\varepsilon^2 p}{\rho U_{\mathrm{ref}}^2}.
\end{equation}
Using the defined dimensionless velocity and length scales, the dimensionless electrolyte-flow equations become
\begin{equation}
\begin{aligned}
\bm{v}^*\cdot\nabla^*\bm{v}^*
&=
-\nabla^*p^*
+
\frac{\varepsilon}{\mathrm{Re}_{D}}\Delta^*\bm{v}^*
-
\frac{\varepsilon^2}{\mathrm{Re}_{K}}\bm{v}^*
-
\varepsilon^2
\frac{c_F}{\sqrt{\mathrm{Da}_{K}}}
|\bm{v}^*|\bm{v}^*,
\\
\nabla^*\cdot\bm{v}^*
&=0.
\end{aligned}
\end{equation}
The dimensionless numbers appearing in the electrolyte-flow equation are
\begin{equation}
\mathrm{Re}_{D}
=
\frac{\rho U_{\mathrm{ref}}L_{\mathrm{ref}}}{\mu},
\qquad
\mathrm{Da}_{K}
=
\frac{K}{L_{\mathrm{ref}}^2},
\qquad
\mathrm{Re}_{K}
=
\mathrm{Re}_{D}\mathrm{Da}_{K}
=
\frac{\rho U_{\mathrm{ref}}K}{\mu L_{\mathrm{ref}}},
\qquad
c_F
=
\beta\sqrt{K}.
\end{equation}

\section{Uniform-concentration charge-conservation limit at high flow velocity}
\label{app:specific-regime}

For a fixed imposed current density, as in galvanostatic operation, when the velocity scale \(U_{\mathrm{ref}}\) is very large, the Péclet number becomes large,
$\mathrm{Pe}
=
\frac{U_{\mathrm{ref}}L_{\mathrm{ref}}}{D_{\mathrm{ref}}}
\gg 1,$
while the convective Damköhler number becomes small,
$\mathrm{Da}_c
=
\frac{a i_{\mathrm{ref}}L_{\mathrm{ref}}}{F c_{\mathrm{ref}}U_{\mathrm{ref}}}
\ll 1$.
Therefore, the diffusion, electromigration, and Faradaic source terms are small compared with advection. The steady-state form of the dimensionless species mass-balance equation~\eqref{eq:dimensionless_species_balance} reduces to
\begin{equation}
\nabla^*\cdot
\left(
\bm{v}^*c_j^*
\right)
\approx 0.
\label{eq:reduced_UCC_MB}
\end{equation}
If the velocity field is divergence-free, $\nabla^*\cdot\bm{v}^*=0$, then 
\begin{equation}
\bm{v}^*\cdot\nabla^*c_j^*
\approx0.
\label{eq:div_advection}
\end{equation}
We further assume that the inlet concentration of each species is spatially uniform, every relevant streamline originates at the inlet, and concentration boundary layers induced by membrane transport are negligible. 
Under these assumptions, every streamline carries the same inlet concentration throughout the flow-connected domain.
Consequently,
\begin{equation}
    c_j^*(\bm{x})
    \approx
    c_j^{\mathrm{res},*},
    \qquad
    \nabla^*c_j^*
    \approx
    \bm{0}.
    \label{eq:uniform_concentration_limit}
\end{equation}
Thus, for a fixed imposed current density and sufficiently high electrolyte velocity, the residence time is short and reaction-induced concentration changes are negligible. 
The species concentrations can therefore be approximated by their spatially uniform reservoir values, and the species mass-balance equations need not be solved.

\twocolumngrid

\bibliographystyle{apsrev4-2}
\bibliography{references}

\end{document}